\documentclass[11pt]{article}
\usepackage[a4paper,left=2.0cm,right=2.0cm,top=3cm,bottom=3cm]{geometry}
\usepackage{amsmath}
\usepackage{bm}
\usepackage[hidelinks]{hyperref}
\usepackage{natbib}
\usepackage{color}
\definecolor{lightgray}{gray}{0.7}    
\definecolor{darkblue}{rgb}{0,0,0.7}
\definecolor{darkred}{rgb}{0.7,0,0}
\usepackage{soul}
\usepackage[normalem]{ulem}
\usepackage{graphicx}
\usepackage{lineno} 
\usepackage{amsmath,amssymb}
\usepackage{gensymb}
\usepackage[nodisplayskipstretch]{setspace}
\usepackage{caption}
\usepackage[finalnew]{trackpackage} 

\newcommand{\s}[1]{\sigma_{#1}}
\newcommand{\si}[1]{\sigma_{#1}}
\newcommand{\gam}{\left( \dfrac{\gamma - 1}{\gamma + 1}   \right)}

\newcommand{\ff}{\dfrac{4\mu_{0}^{2}}{(\gamma+1)^{2}}}
\newcommand{\avg}{\left(\dfrac{\sigma_{xx}+\sigma_{yy}}{2}\right)}
\newcommand{\radius}{\sqrt{\left(\dfrac{\s{xx}-\s{yy}}{2}\right)^2+\s{xy}^2}}

\usepackage{booktabs}
\usepackage{multirow,bigstrut}
\usepackage{rotating}

\usepackage{titlesec}
\titleformat{\subsubsection}[runin]
{\normalfont\bfseries}{\thesubsubsection}{1em}{}
\begin{document}

\begin{center}
\textbf{\Large {On the importance of 3D stress state in 2D earthquake rupture simulations with off-fault deformation}} \\[20pt]
\textcolor{blue}{\small Manuscript accepted to be published in Geophys. J. Int.}\\[20pt]

Louise Jeandet Ribes$^1$,  Marion Y. Thomas$^1$ and Harsha S. Bhat$^2$ \\[20pt]

\begin{enumerate}
\footnotesize
\it
\itemsep0em
\item{Institut des Sciences de la Terre Paris, Sorbonne Universit\'e, CNRS-UMR 7193, Paris, France.}
\item{Laboratoire de G\'{e}ologie, \'{E}cole Normale Sup\'{e}rieure, CNRS-UMR 8538, PSL Research University, Paris, France.}
\end{enumerate}

\end{center}
\vspace{0.2cm}

\noindent{\bf 
 During the last decades, many numerical models have been developed to investigate the conditions for seismic and aseismic slip. Those models explore the behavior of frictional faults, embedded in either elastic or inelastic medi\change[review]{ums}{a}, and submitted to a far field loading (seismic cycle models), or initial stresses (single dynamic rupture models). Those initial conditions impact both on-fault and off-fault dynamics. Because of the sparsity of direct measurements of fault stresses, modelers have to make assumptions about these initial conditions. To this day, Anderson's theory is the only framework that can be used to link fault generation and reactivation to the three-dimensional stress field. In this work we look at the role of the three dimensional stress field in modelling a 2D strike-slip fault under plane-strain conditions. We show that setting up \change[review]{the}{an} incorrect initial stress field, based on Anderson's theory, can lead to underestimation of the damage zone width by up to a factor of six, for the studied cases. Moreover, because of the interactions between fault slip and off-fault deformation, initial stress field influences the rupture propagation. Our study emphasizes the need to set up the correct initial 3D stress field, even in 2D numerical simulations. }\\



\section{Introduction: Modelling fault slip}
\label{sec:intro}

%

Catastrophic, seismic events of large magnitude  ($M_w > 7$) remain relatively rare, with a recurrence time of several decades up to a millennium \citep{Cubas2015}. As a consequence, observations are sparse and numerical models are powerful tools to explore the conditions for seismic and aseismic fault slip. Numerical models that account for both seismic slip and  long-term slow slip are challenging because of the wide range of temporal and spatial scales involved \citep{lapusta2000elastodynamic}. Hence,  compromises are made to reduce the computational cost depending on the goal of the model. These compromises can be safely categorized as models that focus on a single dynamic rupture event and ones that model \remove[review]{the} multiple seismic cycles.

Single dynamic rupture models were some of the first analytical and numerical models developed for earthquakes. They reproduce an event from the moment it turns dynamic to its arrest and have provided important insights into earthquake mechanics \citep[among many others that followed]{kostrov1964, andrews1976, madariaga1976, das1986a, das1987}.

For seismic cycle models,  the focus is on integrating all stages of fault slip (inter-, co- and post-seismic) over thousands of year\add[review]{s} \citep{Erickson2020}.  
This is indeed critical as pre-stress inherited from aseismic slip and prior seismic events likely play a determinant role on where earthquakes will nucleate and how far their rupture will propagate \citep{thomas2014quasi}. The most popular strategy is to model a single planar fault governed by \change[review]{R\&S}{rate-and-state (R\&S)} friction law, embedded in a purely elastic medium. \citep[e.g.,][]{Ben-Zion1997,lapusta2000elastodynamic,RichardsDinger2000, Kato2004,Barbot2012,Im2020,Liu2020}. In these models, frictional heterogeneities are invariably advocated to reproduce the full slip spectrum (creep, slow slip events, earthquakes, etc...). More recent models also account for complex fault geometry \citep[among others]{romanet2018fast, ozawa2021, Uphoff2022}. However, due to timescales that vary over several decades of orders \add[review]{of magnitude}, the most popular compromise that is made is to ignore the wave mediated stress transfer (inertial dynamics) and only account for linear elastic static stress transfer along with the instantaneous, local traction contribution. Such class of models are called quasi-dynamic models.

Natural fault zones, however, i.e. the fault plane and its surrounding medium, are intricate structures in constant evolution in response to tectonic loading. As an example, during an earthquake, part of the stored elastic strain energy is dissipated in  off-fault deformation, or damage, which in turn radiate and affect the slip dynamics of the main fault. 
If we provide an idealized description, fault zones comprise of a non-planar fault core,  where most of the displacement has occurred, surrounded by a damage zone that has a spatial scale of the order of meters to kilometers \citep[e.g.,][]{Chester1993,Biegel2004,Faulkner2011}.  Frequently,  the fault core corresponds to an extremely narrow band and the damaged wall rock includes layers of gouge and breccia bordered by fractured rocks. The last two layers are included in the damage zone because they lacked extensive shearing. 
These structures are of key importance in the mechanics of faulting. For example, fault roughness has an effect on the fault resistance to slip, i.e., the fault strength \citep[e.g.,][]{Dunham2011b,Tal2022}. Laboratory experiments of earthquakes in a damaged medium show that there is an intimate interaction between the rupture and off-fault damage zone \citep{sammis2009,bhat2010a,biegel2010}. The density of this damage has a direct impact on the elastic moduli of the bulk \citep{Walsh1965a,Walsh1965,faulkner2006slip}, therefore, on the quantity of strain energy which is stored and further released by fault slip.
In fact, systematic micro- and macrostructural field studies have been performed on damage zones \citep[e.g.,][]{Shipton2001,manighetti2001slip,manighetti2004role,Dor2006,mitchell2009nature,Savage2011,Johnson2021,Rodriguez-Padilla2022}, as a key component to understand the energy balance of earthquakes \citep[e.g.,][]{Rice2002,Kanamori2006,okubo2019dynamics}.
Off fault damage is observed from the mesososcopic scale to the microscopic scale, with a microfracture density that decreases exponentially away from the fault core  \citep{mitchell2009nature}. The width of the damage zone is also believed to be decreasing with depth,  forming a ``flower-like structure” \citep{ben2003shallow,cochran2009seismic}. However  \citet{okubo2019dynamics} have demonstrated numerically that, despite its reduction in spatial-extent with depth, energetically speaking, the contribution of the off-fault damage increases with depth.

Thus,\add[review]{as oppose to seismic cycle codes}, a second set of models have been developed to catch the full slip dynamics, the wave propagation and the interactions with the off-fault medium during an earthquake.  With these models, \change[review]{people}{researchers} have explored the effect of complex fault geometry and/or the effect of off-fault damage on seismic rupture.  Some models treat the bulk as a solid linear-elastic material and prescribe a low-velocity zone around the fault to account for damage \citep[e.g.,][]{cappa2014off,Huang2014a}.  Another set of models \change[review]{have}{has} explored the effect of spontaneous dynamic generation of off-fault deformation using a Mohr-Coulomb \citep[e.g.][]{andrews2005rupture,ben2005dynamic,Hok2010,Gabriel2013} or Drucker-Prager \citep[e.g.][]{templeton2008off,Ma2008,Dunham2011b,Johri2014} based plastic constitutive laws. Another class of models have treated off-fault damage as tensile cracks, using a stress- \citep{yamashita2000generation} or fracture-energy-based \citep{dalguer2003simulation} criterion. \citet{okubo2019dynamics} used a Finite Discrete Element Method (FDEM) that allows spontaneous nucleation and propagation of off-fault fracture network in a medium. These studies have provided a good insight on the effect of a fault zone structure on dynamic ruptures but the models do not account for the  observed coseismic change of elastic properties in the bulk \citep{hiramatsu2005seismological,Brenguier2008,froment2014imaging} which also impacts the rupture dynamics. This can be achieved by using  \change[review]{an}{a} homogenized damage mechanics based constitutive law\remove[review]{s} \citep[e.g.,][]{Lyakhovsky1997,bhat2012,Xu2014,thomas2017effect,thomas2018dynamic}. A vast majority of the investigations cited above are done under two-dimensional plane-strain conditions.

Finally, some models have been developed to look at the zeroth order effect of the fault core (as opposed to a fault interface) or the damage zone on the seismic cycle.  To overcome the computational cost,  they have to compromise on both the fault slip dynamics and the dynamics of bulk evolution (off-fault crack growth). 
As an example,  \citet{Ende2018} have represented the fault core by a shear band but the bulk remains elastic.
\citet[][]{Kaneko2011,Idini2020}, and \citet{Abdelmeguid2019} among others, looked at the effect of a prescribed low velocity fault zones (LVFZ), but, by construction, the bulk is still a passive elastic body.  They used quasi-static or quasi-dynamic approximations to solve the problem.  In their model, \citet{Thakur2021} imposed (thus not driven by the model) a time-dependent shear modulus evolution to account for coseismic drop and postseismic recovery of elastic moduli.  \citet{Erickson2017} has applied quasi-dynamic analysis to explore the effect of plasticity throughout the earthquake cycle.  \citet{Preuss2020} have developed a 2.5D model with a visco-elasto-plastic crust subjected to rate- and state-dependent friction to model conjointly the rapid deformation in the elastic–brittle upper crust and the relaxation in the deeper viscoelastic crustal substrate and their influence on each other\remove[review]{s}.

In all cases,  irregardless of the end goal, setting the initial and boundary conditions (initial stresses for a single rupture, far field loading for seismic cycles) directly impacts both on-fault and off-fault dynamics. 
But what constraints do modellers have on the stress state of a fault?
This will be discussed in the section below, followed-up by a review on how boundary conditions are set up in the community. We will then present a simple method to accurately define the initial stress field for in-plane conditions. 
To demonstrate its importance, we will explore its influence on\remove[review]{ off}-fault and off-fault deformation.  Results are summarized and discussed in the last section.

\section{Stress field and faulting}
\label{sec:stressandfault}
\subsection{The Anderson theory}
\label{subsec:andersontheory}

\begin{figure*}
\centering
\includegraphics[width=1\textwidth]{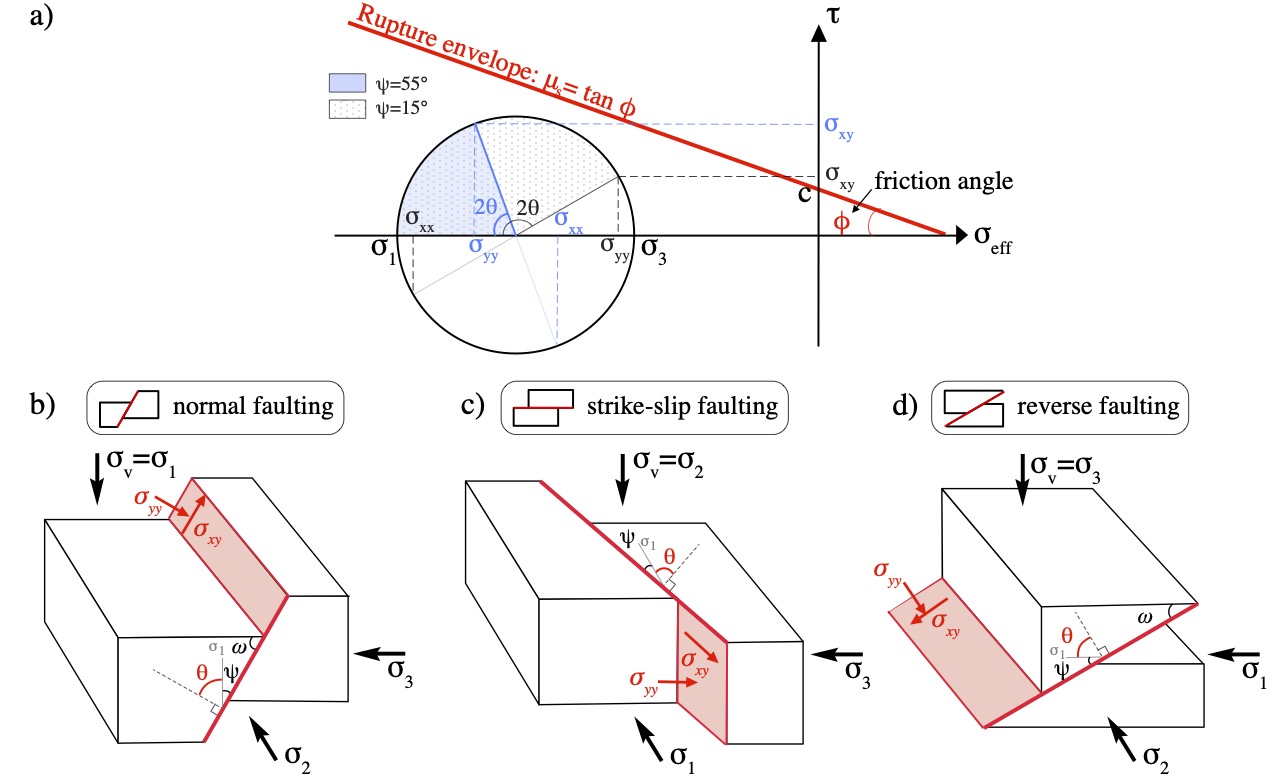}
\caption{a) Mohr–Coulomb criterion in ($\sigma,\tau$)-space.   b), c) \& d) Orientation of the failure plane with respect to the largest principal stress for normal, strike-slip, and reverse faulting respectively. \add[review]{The angle $\phi$ correspond to the slope of the failure envelope on a Mohr diagram and defines the static friction $\mu = \tan\phi$. The remaining angles $\Psi$, $\theta$ and $\omega$ are defined within the plane perpendicular to $\sigma_2$.   $\Psi$ and $\theta$ corresponds to the angle between $\sigma_1$ and the fault plane ($\Psi$) or,  following the Mohr-Coulomb convention, the normal to the fault plane ( $\theta$).  $\omega$ gives to the dip angle of the fault.}}
\label{fig01orientation}
\end{figure*}

In the upper crust, large strains are accommodated by fault systems either seismicaly or aseismically. It either leads to the formation of new fractures in the crust, or reactivates previously existing faults. Fault systems evolve and acquire their general geometry by the progressive amalgamation of such fractures \citep{cowie1992growth}. 
In-situ measures of stresses on a fault, at any location, at any time are impossible to achieve, hence it has to be approached theoretically.
A connection between the geometry of fault systems and the forces that formed them was first proposed by \citep{anderson1905dynamics}. His theory related the initial formation of faults to the state of stress in the crust, under the assumption that rocks are isotropic, homogeneous and intact. 
This theory is based on the mathematical result that at every point of a stressed rock, three planes can be found on which no shear traction\remove[review]{s} acts. Those planes are perpendicular \change[review]{one to}{to one} another and are called the principal planes. The corresponding stresses acting along the three principal directions are called the principal stresses. \change[review]{(stresses positive in tension)}{By convention, the stresses are assumed to be positive in tension. Therefore, in a compressive regime, we have the following inequality : }
\begin{linenomath*}\begin{equation}
\s{1} < \s{2} < \s{3}
\label{eqn:pal_stress}
\end{equation} 
\end{linenomath*}
Then Anderson assumed that, apart from those three planes,  a plane with maximum tangential stress exists, on which the faulting should initiate. 
For brittle shear failure, and for a fluid saturated rock mass with pore fluid pressure $P_f$ \add[review]{(with $P_f > 0$)}, this Coulomb failure criterion may be written as: 
\note[review]{added negative sign for $\sigma_{eff}$ and $\sigma_{n}$}{
\begin{linenomath*}\begin{equation}
\tau_y = c + \mu(-\sigma_{eff})= c + \mu(-\sigma_n - P_f)
\label{eqn:coulomb_failure}
\end{equation} \end{linenomath*}}

where $\sigma_n$ is the normal stresses acting on the plane and \note[review]{changing sign for pore pressure}{$\sigma_{eff}=\sigma_n + P_f$} corresponds to the effective normal stress. The variable $c$ is the cohesion and $\mu = \tan\phi$ is the coefficient of friction.  The angle $\phi$ corresponds to the slope of the failure envelope on a Mohr diagram \add[review]{(Figure}~\ref{fig01orientation}) and $\theta$, \add[review]{on the same diagram}, corresponds to  the angle between the normal to the fault plane and $\sigma_1$ (Figure~\ref{fig01orientation}). Hence we can write:
\begin{linenomath*}\begin{equation}
  \begin{array}{ll}
\s{yy} =& \left(\dfrac{\s{1}+\s{3}}{2}\right)- \left(\dfrac{\s{3}-\s{1}}{2}\right)\cos{2\theta} \\
\s{xx} = &  \left(\dfrac{\s{1}+\s{3}}{2}\right)+ \left(\dfrac{\s{3}-\s{1}}{2}\right)\cos{2\theta}\\
\s{xy} = & \left(\dfrac{\s{3}-\s{1}}{2}\right)\sin{2\theta} \\
 \end{array}   
\label{eqn:normalshearstress}
\end{equation} 
\end{linenomath*}
%
%
%
The radius $R$ of the Mohr circle is then given by:
\begin{linenomath*}\begin{equation}
R=\left(\dfrac{\s{3}-\s{1}}{2}\right)=\radius
\label{eqn:raidus}
\end{equation} \end{linenomath*}
%

In numerical studies, it is more common to define $\Psi$, the angle between the fault plane and  $\sigma_1$ (Figure~\ref{fig01orientation} \& \ref{fig02setup}), such as $\Psi=\pi/2-\theta$.  
Hence, as  illustrated in Figure~\ref{fig01orientation}a, the relationship between the magnitudes of $\s{yy}$ and $\s{xx}$ depends on $\Psi$ in the following manner
\begin{linenomath*}\begin{equation}
\begin{array}{ccc}
|\s{yy}| \leq  |\s{xx}| & \mbox{if} & 0\degree \leq\Psi \leq 45\degree \\
|\s{yy}| > |\s{xx}| & \mbox{if} & 45\degree < \Psi \leq 90\degree \
\end{array}
\label{eqn:SyySzz}
\end{equation}\end{linenomath*}

Then, for an optimal angle $\theta_{opt}=\pi/4 +\phi/2$ shear failure occurs on a plane containing the $\sigma_2$ direction, when $\s{yy} =\s{eff}$ and  $\s{xy} =\tau_y$, i.e., when the failure envelop is tangential to the Mohr's circle.
Relying on experimental studies, the internal friction $\mu$ generally lies between 0.5 and 1 (Jaeger and Cook, 1979). Hence the optimal angle $\theta_{opt}$ varies between 58$\degree$  and 68$\degree$ and the new fault should make an angle $\Psi$ with $\sigma_1$  between 32$\degree$ and 22$\degree$ respectively.  
Following the same reasoning for a pre-existing fault plane,  if we consider a static coefficient of friction $\mu_s=0.6$, which corresponds to a large variety of rocks and minerals \citep{byerlee1978friction},  fault\add[review]{s} are optimally oriented when they make\remove[review]{s} an angle $\Psi\simeq30\degree$ with $\sigma_1$. 
 
\change[review]{Anderson theory is based on the assumption that Earth's free surface requires a principal stress direction to be subvertical ($\sigma_v$).}{The foundation of the Anderson's theory lies in the observation that, due to Earth's free surface, it is imperative for one principal stress to be oriented subvertically, i.e., to be equal to $\sigma_v$.}
This gives rise to three fundamental stress regimes (Figure \ref{fig01orientation}) depending on whether $\sigma_v=\sigma_1$, $\sigma_2$ or $\sigma_3$: reverse faulting ($\sigma_v = \sigma_3 $),  strike-slip faulting ($\sigma_v = \sigma_2 $), and normal faulting ($\sigma_v = \sigma_1 $).  Keeping $\mu_s=0.6$ as a reference value for the static friction,  it is then expected to get sub-vertical strike-faults  and a dip angle of $\omega \approx 30\degree$ and $\omega \approx 60\degree$ for a thrust and a normal fault respectively (Figure \ref{fig01orientation}).

Despite the simplicity of the theory, observations are in accordance with the model for strike slip faults with low cumulative displacement \citep{anderson1951dynamics,kelly1998linkage}. Earthquakes have also been registered on subvertical fault striking approximately 30$\degree$ to the regional $\sigma_1$ with a subvertical $\sigma_2$: the 2000 Western Tottori earthquake in Japan \citep{sibson2012andersonian,fukuyama2003detailed,yukutake2007estimation} or the 2010 \add[review]{Mw 7.1} Darfield earthquake \remove[review]{Mw 7.1}  in New Zealand \citep{sibson2012andersonian}. Moreover, borehole measurements and induced seismicity \citep{townend2000faulting}, paleostresses inversion \citep{lisle2006favoured} and earthquake focal mechanisms \citep{celerier2008seeking} suggest that Andersonian state of stress prevails over large areas within the shallow crust. 
Of course exceptions to the theory exist too. Well-known examples, such as the San Andreas fault \citep{Zoback1987}, or the low-angle normal faults in  Elba, Central Italy \citep{Collettini2001} or in the Cyclades Greece \citep{Lecomte2010} are mis-oriented if we consider a coefficient of friction of 0.6 \citep{byerlee1978friction}.  But a lower friction on the fault plane (clay minerals) or a high pore pressure may well explain the discrepancy.  If the fault essentially slips during earthquakes,  weakening mechanisms such as the ones listed by \citet{Tullis2015} may well kick in (mineral breakdown, flash melting, thermal pressurization, etc...), leading to a very low effective coefficient of friction.
\add[review]{Finally, polymodal faulting, with three or more sets of faults forming and slipping simultaneously are not compatible with the Anderson's assumption that faults form parallel to the intermediate principal stress, $\sigma_2$ } \citep{Healy2015}.
Hence, \change[review]{including}{excluding} these exceptions, it suggests that the Anderson\add[review]{'s} theory provides a useful framework for general considerations about fault generation and reactivation.  And considering the lack of systematic, time-dependent in-situ stress measurements, probably the only one.

\subsection{Non-exhaustive review on how initial stress state \change[review]{in}{is} prescribed in numerical models}
\label{subsec:BC}

In \change[review]{2D}{single dynamic rupture} models, rupture grows under the control of the prescribed initial stresses. Theoretical analyses \citep{poliakov2002dynamic,rice2005off,ngo2012off} have illustrated the effect of initial stress field on the pattern of off fault damage activation (i.e., the potential failure area, and the orientation of secondary cracks) around a propagating crack. 
They show that the extent and location of secondary faulting (the activated zone) is strongly affected by the orientation of principal stresses, set up by $\Psi$ (the angle between $\sigma_1$ and the primary fault). Steep $\Psi$ favors inelasticity on the extensional side and shallow $\Psi$ on the compressional side. Moreover initial stresses \remove[review]{seems} seem to influence the potential for rupture to follow intersecting faults with different orientation rather than the primary fault \citep{kame2003b, bhat2004, fliss2005}. Therefore, pre-stress orientation is a key parameter in numerical simulation of dynamic rupture with off-fault inelastic deformation.  

Most of the numerical studies investigating the interactions between seismic rupture propagation and  off fault damage usually set up a 2D planar strike-slip fault under plane-strain conditions \citep{andrews2005rupture,Shi2006,templeton2008off,dunham2011earthquake,thomas2017effect, okubo2019dynamics} , such as displayed in Figure~\ref{fig01orientation}. 
%
Because of the 2D setting, the out of plane stress is often ignored when setting up the initial stress field. Usually, a normal stress and a shear stress are imposed, corresponding to two principal horizontal stresses $\sigma_1$ and $\sigma_3$ and an angle $\Psi$ between fault and main stress direction. However, whether a fault is under a strike-slip stress field depends on the full 3D stress field (Figure~\ref{fig01orientation}). If the two in-plane stresses are $\sigma_1$ and $\sigma_3$, then the fault is in a proper strike-slip stress field. If the two in-plane stresses are $\sigma_1$ and $\sigma_2$, then the fault is in a reverse stress field. This would not change the slip on the fault because slip is restricted to the 2D plane. However, the ratio between the out of plane stress and the smallest principal in\add[review]{-}plane stress would be \change[review]{greater}{smaller} than expected under a proper 3D stress strike-slip stress field. 

Despite the importance of initial stress field in modeling off fault deformation, the \textit{full} three dimensional initial stress field and the importance of the out-of-plane stress, in 2D simulations, have never been discussed. 
\subsection{Criterion for accurate initial stress field in 2D plane-strain simulations}
\label{subsec:criterion}

\begin{figure}
\centering
\includegraphics[width=1\textwidth]{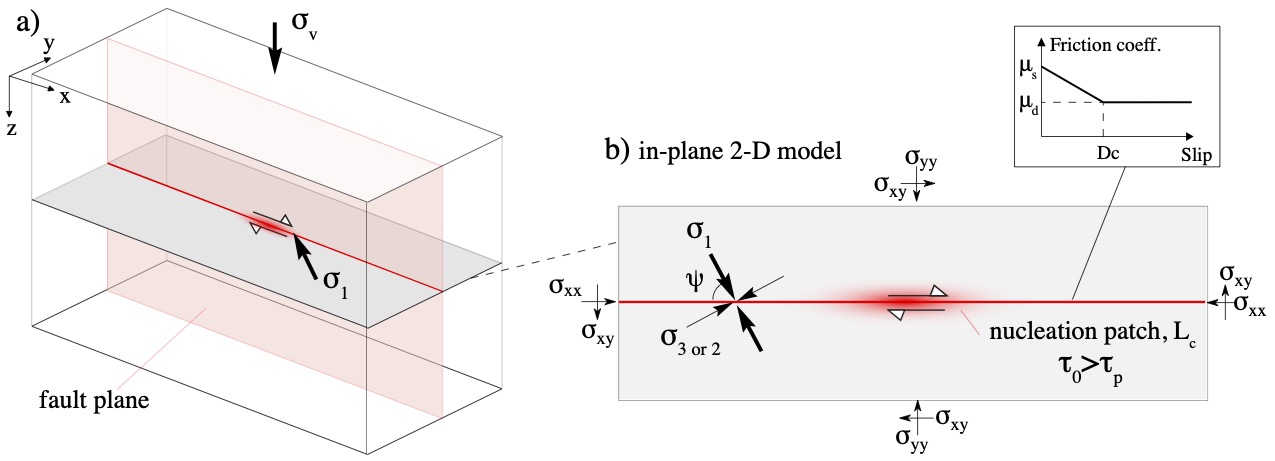}
\caption{Modeling set up for a dynamic rupture on a strike-slip fault. a) Schematic of the fault zone in 3D.  b) Zoom on the 2D plane that hosts the modeled rutpure.  The orientation of maximum compressive stress $\sigma_1$ is set at an angle $\Psi$ to the fault plane that is governed by slip weakening friction law. The nucleation patch is set by increasing the initial shear stress $\si{xy}$ slightly above the fault strength $\mu_s\left(-\si{yy}\right)$ over a length $L_c$.}
\label{fig02setup}
\end{figure}

When considering a planar, strike-slip fault under plane-strain condition (Figure \ref{fig02setup}). The initial stress state, $\si{ij}$, (tensile positive) is represented as,
\begin{linenomath*}\begin{equation}
\boldsymbol{\bar{\sigma}}=\si{ij} \equiv
\begin{pmatrix}
\s{xx} & \s{xy} & 0 \\ 
\s{xy} & \s{yy} & 0 \\ 
0 & 0 & \s{zz}  
\end{pmatrix}
\label{eqn:stressmatrix}
\end{equation}\end{linenomath*}
For convenience, we define the following ratios:
\begin{linenomath*}\begin{equation}
\gamma = \dfrac{\s{xx}}{\s{yy}} \textrm{  ~\&~  } \mu_{0} = \dfrac{\s{xy}}{-\s{yy}}
\label{eqn:ratio}
\end{equation}\end{linenomath*}
 
which leads to:
\begin{linenomath*}\begin{equation}
\label{eqn:Psi}
        \tan{2\Psi} =-\tan{2\theta}= -\dfrac{2\s{xy}}{\s{xx}-\s{yy}}=\dfrac{2\mu_0}{\gamma-1}\\
\end{equation}\end{linenomath*}

Let $\s{1},\s{2} $ and $ \s{3}$  represent the maximum, intermediate and minimum compressive principal stresses.  For a strike-slip fault, $\s{1} $ and $ \s{3}$ should lie on the $x-y$ plane and $\s{2}$ should be parallel to the $z$ axis, i.e out of plane for a 2D simulation (Figure~\ref{fig01orientation}). Hence, 
\begin{linenomath*}\begin{equation}
\label{eqn:principalstress}
    \begin{array}{l}
        \s{1} = \avg-R=\left(\dfrac{\s{xx}+\s{yy}}{2}\right)\left(1 + \sqrt{\gam^{2} + \ff}   \right),\\
	    \s{3} = \avg+R=\left(\dfrac{\s{xx}+\s{yy}}{2}\right)\left(1 - \sqrt{\gam^{2} + \ff}   \right).\\
        \s{2} = \s{zz} = \rho g z (1-\lambda)\\
    \end{array}
\end{equation}\end{linenomath*}
where $\lambda$ is the pore pressure coefficient \add[review]{ and $z$ is a parameter corresponding to the depth of the 2D slice where plane strain simulations are conducted}.
Under the plane-strain approximation, we also have the following relationship:
\begin{linenomath*}\begin{equation}
\s{2} = \s{zz} = \nu(\s{xx}+\s{yy}) = \nu \s{yy} (\gamma+1) 
\label{eqn:out_of_plane_stress}
\end{equation}\end{linenomath*}
where $\nu$ is the Poisson's ratio. 
%
%
%
%

\begin{figure}
\centering
\includegraphics[width=1\textwidth]{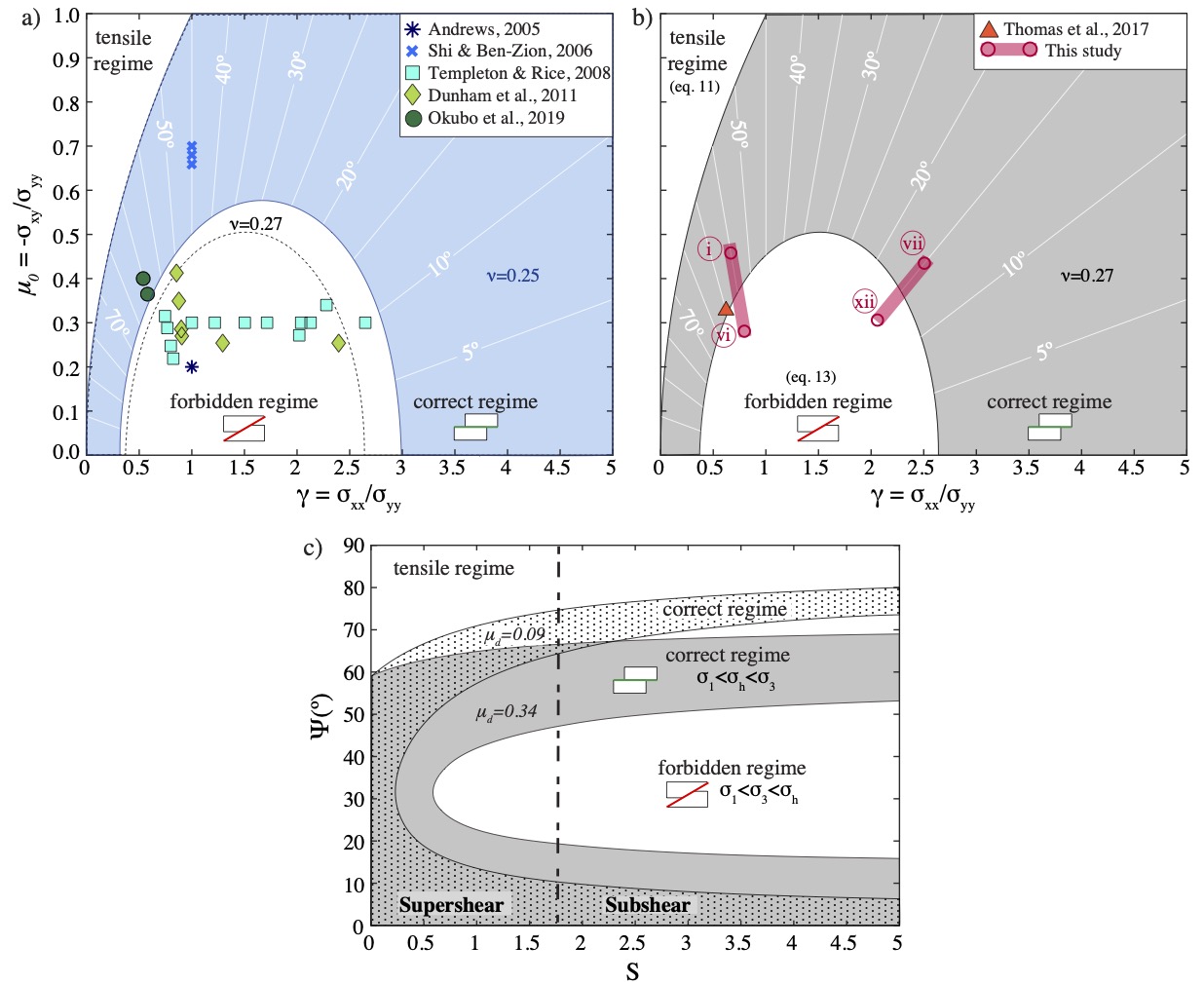}
\caption{Criterion for accurate initial stress field under plane-strain approximation displayed as $\gamma-\mu_0$, plots (a) \& (b) and $\Psi-S$ for plot (c). 
The \change[review]{gray}{shaded} areas represent the conditions for which the initial stresses favor a strike-slip motion, i.e. when equation (\ref{eqn:criterion}) is satisfied. 
In the $\gamma-\mu_0$ plot, this criterion \add[review]{depends on $\nu$ (0.25 and 0.27 for plot (a) and (b) respectively), but} is independent from other parameters. Contours lines shows $\Psi$ values. Color dots shows the initial parameters for published simulations of strike-slip faulting with plane-strain approximation. \add[review]{In plot (b), the} shaded red areas shows the two sets of parameters explored in this study, with  $\Psi = 15^o$ and $\Psi = 55^o$. The red empty circles correspond to the end members.  In subplot (c) the criterion is represented as a function of $S$ and $\Psi$ for $ \mu_s=0.6$. Plotting the area corresponding to an accurate initial stress depends on a third parameter, here given by $\mu_d$.  We plot the criterion for two cases, $\mu_d$ = 0.09 (dotted area) and $\mu_d$ = 0.34 (gray area).}
\label{fig03condition}
\end{figure}

We require that both $\s{1}$ and $\s{3}$ should be compressive, i. e. $\s{1}<\s{3}<0$. This implies that
\begin{linenomath*}\begin{equation}
{\sqrt{\gam^{2} + \ff}  ~<~ 1}
\label{eqn:pal_stress_neg2}
\end{equation}\end{linenomath*}
since $\s{xx}+\s{yy}<0$. The principal stress state must also satisfy the inequality given in equation~\ref{eqn:pal_stress}.
\begin{linenomath*}\begin{equation}
\label{eqn:s1s2}
    \begin{array}{ll}
 	   \s{1} < \s{2} & \Rightarrow \left(\dfrac{\s{xx}+\s{yy}}{2}\right)\left(1 + \sqrt{\gam^{2} + \ff}   \right) ~<~ \nu(\s{xx}+\s{yy})\\
        & \Rightarrow {\sqrt{\gam^{2} + \ff}  ~>~ 2\nu - 1}
     \end{array}
\end{equation}\end{linenomath*}
\add[review]{ This is trivially satisfied as $\nu$ is always smaller than 0.5.}
\begin{linenomath*}\begin{equation}
\label{eqn:s2s3}
    \begin{array}{ll}
 	   \s{3} > \s{2} &\Rightarrow \left(\dfrac{\s{xx}+\s{yy}}{2}\right)\left(1 - \sqrt{\gam^{2} + \ff}   \right) ~>~ \nu(\s{xx}+\s{yy}) \\
                           &\Rightarrow {\sqrt{\gam^{2} + \ff}  ~>~ 1 - 2\nu}
     \end{array}
\end{equation}\end{linenomath*}
Thus, the stress field for a strike-slip fault must satisfy the following criterion:
\begin{linenomath*}\begin{equation}
{1-2\nu ~<~ \sqrt{\gam^{2} + \ff}  ~<~ 1}
\label{eqn:criterion}
\end{equation}\end{linenomath*}
This inequality is plotted in Figure \ref{fig03condition} as a function of $\gamma$ and $\mu_0$.  The \change[review]{area in grey}{shaded areas} define what we will from now on refer to as the ``correct'' regime (strike-slip stress field). The white areas represent (1) the ``forbidden'' regime (reverse faulting stress field) for which, this criterion is violated and (2) the stress field corresponding to a tensile regime. Superimposed on this graph are the initial parameters for several studies modeling strike-slip motion using plane-strain approximation. Three out of six studies used a far field stress field that favors reverse faulting,  i.e the out-of-plane stress is $\s{3}$ and not $\s{2}$ as it should be (Figure \ref{fig03condition}a). 
If the modelling is only performed on a 2D plane,  like the vast majority of the published studies,  the fault will still have a strike-slip motion even if this condition is not satisfied simply because the motion is restricted along one plane.  However,  the whole stress field would favor reverse faulting, which will impact any model of off-fault deformation as we will demonstrate soon.

In Figure \ref{fig03condition}c, we plot the same criterion as a function of $\Psi$ and the seismic ratio $S$, as defined by \citep{andrews1976,das1977},
\begin{linenomath*}\begin{equation}
S = \dfrac{\mu_s(-\s{yy}) -\s{xy} }{\s{xy} - \mu_d(-\s{yy}) } = \dfrac{\mu_s- \mu_0 }{\mu_0 - \mu_d }
\label{eqn:Sratio}
\end{equation}\end{linenomath*}
where $\mu_s$ and $\mu_d$ correspond to the static and dynamic coefficient of friction respectively,  for a slip weakening law.  $S$ helps determine whether a rupture in 2-D, \add[review]{under homogeneous conditions,} can reach supershear velocities (S $<$ 1.77), or remains sub-Rayleigh (S $>$ 1.77). 
For a fixed value of $\Psi$,  fulfilling the criterion for strike-slip faulting strongly depends on $\mu_d$.  
Interestingly, setting up a proper strike-slip stress field in agreement with Anderson theory, i.e. $\Psi \simeq 30^\circ$  for $\mu_s=0.6$, requires supershear parameters,  and this holds true for a large range of realistic $\mu_d$ values.
%

\section{Methods}
\label{sec:method}

%
\subsection{Numerical model setup}
\label{subsec:setup}

In this study, we explore the boundary between the forbidden regime and the correct regime by modeling rupture on a 1D right-lateral fault in a 2D medium under plane-strain approximation (Figure \ref{fig02setup}). We particularly focus on the influence of the \change[review]{pre-stress}{initial stress} conditions on off-fault stresses and inelastic deformation.  
We use the 2-D spectral element code SEM2DPACK \citep{ampuero2012sem2dpack}.
Rupture propagation along the fault plane is governed by a slip-weakening friction law \citep[e.g., ][]{palmer1973growth}. Slip occurs when the on-fault shear stress reaches the shear strength $\tau_f= \mu^*(-\s{yy})$. The friction coefficient $\mu^*$ depends on the cumulated slip ($\delta$) and drops from a static  $\mu_s$  to a dynamic $\mu_d$ value over a characteristic distance ($D_c$). 
The rupture is artificially nucleated within a patch where the initial shear stress $\s{xy}$ is set just above the fault strength (Figure~\ref{fig02setup}). Further on, since the difference in nucleation duration $t_{nuc}$ between two models has no physical meaning, we will shift all the results in time so $t_{nuc}$ = $t_{0}$. 
Following \citet{Kame2003}, the minimum nucleation size $L_c$ determined by the energy balance for a slip weakening law is:
\begin{linenomath*}\begin{equation}
L_c = \frac{16}{3\pi} \frac{\mu G}{(\s{xy} - \tau_r)^2}
\label{eqn:Lc}
\end{equation}\end{linenomath*}
where $G$ is the fracture energy, defined as:
\begin{linenomath*}\begin{equation}
G = \frac{1}{2}D_c (\tau_p - \tau_r)
\label{eqn:fractureenergy}
\end{equation}\end{linenomath*}
and where $\tau_p=\mu_s(-\s{eff})$ and $\tau_r=\mu_d(-\s{eff})$ are respectively the peak and residual stresses.

We further use the process zone size  $R_0$ for a quasi-stationary crack, to normalize the length scales in our results. It corresponds to the length over which the friction drops, with ongoing slip, from the peak strength to the residual strength.  Following \citet{Day2005} it is given by, 
\begin{linenomath*}\begin{equation}
R_0 = \frac{9\pi}{32(1-\nu)} \frac{D_c \mu}{(\mu_s - \mu_d)(-\sigma_{yy})}
\label{eqn:R0}
\end{equation}\end{linenomath*}

In our models, $R_0$ is approximately equal to 1.0 km for $\Psi=55\degree$ and 1.6 km for $\Psi=15\degree$ (see section~\ref{subsec:param} for an explanation on the parameters). We use a resolution of 30 m to ensure that the problem is correctly resolved. In order to scale the problem, all the modelled faults have a length of 30$R_0$.

\subsection{Initial parameters}
\label{subsec:param}
In order to compare our study to the available literature, we run two sets of models with  $\Psi=55\degree$ and $\Psi=15\degree$ respectively (Figure~\ref{fig03condition}a).  To explore the effect of initial stresses on off-fault and on-fault deformation, for each set of models, we run three simulations with the correct strike-slip set-up (cases \textit{i, ii, iii} for $\Psi=55\degree$ and cases  \textit{vii, viii, ix} for $\Psi=15\degree$) and three within the so-called forbidden regime (cases \textit{iv, v, vi} for $\Psi=55\degree$ and cases  \textit{x, xi, xii} for $\Psi=15\degree$).

\begin{table}
\caption{Constants used in all models}
\begin{center}
\begin{tabular}{l cc}
\hline
Parameter & Symbol & Value \\
\hline
depth &$z $& -2.5 km  \\
material density &$\rho$ & 2700 kg.m$^{-3}$\\
S-wave velocity &$c_s$ &  3120 km.s$^{-1}$\\
P-wave velocity &$c_p$ & 5600  km.s$^{-1}$\\
Poisson's ratio &$\nu$ & 0.27 \\
characteristic slip &$D_c$ & 1 m\\
pore pressure coefficient &$\lambda$ & 0.4\\
\hline
\end{tabular}
\label{tab:constants}
\end{center}
\end{table}

\add[review]{To achieve that goal, we adopt the following strategy. }
Some parameters are kept constant between all the simulations: the elastic properties, the depth, the characteristic slip in the friction law and a hydrostatic pore pressure condition (Table \ref{tab:constants}). 
\change[review]{Then, to set up the remaining parameters we adopt the following strategy. First} {We have two $\Psi$ that define the  orientation of maximum compressive stress $\sigma_1$ with respect to the fault.} We set the S ratio (equation~\ref{eqn:Sratio}),  favouring a value that would lead to subshear rupture ($S=2$) for $\Psi = 55\degree$. However, for $\Psi=15$\add[review]{$\degree$}, this lead to a small dynamic stress drop which prevents the rupture from propagating (dying cracks). Hence we choose to set the ratio to $S=1$  for $\Psi=15$\add[review]{$\degree$} (the rupture can evolve to a supershear earthquake). We further fix the stress drop ($\sim10$ MPa):
\begin{linenomath*}\begin{equation}
\Delta\tau = (\mu_0 - \mu_d)(-\s{yy})
\label{eqn:stressdrop}
\end{equation}\end{linenomath*}
and, in agreement with laboratory values \citep{jaeger1979cook}, we make $\mu_s$ varies between 0.52 and 0.68 to get the six models defined above. 

The others parameters can be determined using the following set of equations. In hydrostatic pore pressure condition, the vertical, out-of-plane stress $\s{zz}$ is given by 
\begin{linenomath*}\begin{equation}
\s{zz} = \rho g z (1 - \lambda)
\label{eqn:Szz}
\end{equation}\end{linenomath*}
We then need to compute the ratio $\gamma$:
\begin{linenomath*}\begin{equation}
    \begin{array}{ll}
 	   \text{eq.~\ref{eqn:ratio} \& \ref{eqn:out_of_plane_stress}}   &\Rightarrow \gamma=\dfrac{\s{zz}}{\nu\s{yy}}-1 \\[10pt]
       \text{eq.~\ref{eqn:stressdrop} \& \ref{eqn:Sratio}}   &\Rightarrow \gamma=\dfrac{\s{zz}\left(\mu_0-\mu_s\right)}{\nu S \Delta\tau}-1 \\[10pt]
       \text{eq.~\ref{eqn:Psi}}   &\Rightarrow \gamma=\dfrac{\s{zz}\left(0.5\left(\gamma-1\right)\tan{2\Psi}-\mu_s\right)}{\nu S \Delta\tau}-1 \\[10pt]
        &\Rightarrow \gamma=\dfrac{\left(\alpha+\beta\right)+\mu_s\s{zz}}{\left(\beta-\alpha\right)}\\[10pt]
      & \text{with } \alpha=\nu S\Delta\tau  \text{ \& } \beta = 0.5\s{zz}\tan{2\Psi}
     \end{array}
\end{equation}\end{linenomath*}
\label{eqn:gamma}
Knowing $\gamma$, we can derive:
\begin{linenomath*}\begin{equation}
\mu_{0} =\dfrac{\left(\gamma-1\right)}{2}\tan{2\Psi}
\label{eqn:mu0}
\end{equation}\end{linenomath*}
\begin{linenomath*}\begin{equation}
\mu_{d} =\mu_{0}-\dfrac{\mu_{s}-\mu_{0}}{S}
\label{eqn:mud}
\end{equation}\end{linenomath*}
\begin{linenomath*}\begin{equation}
\s{yy} =\dfrac{\s{zz}}{\nu\left(\gamma-1\right)}
\label{eqn:syy}
\end{equation}\end{linenomath*}
\begin{linenomath*}\begin{equation}
\s{xx} =\gamma\s{yy}
\label{eqn:sxx}
\end{equation}\end{linenomath*}
\begin{linenomath*}\begin{equation}
\s{xy} =\mu_{0}(-\s{yy})
\label{eqn:sxy}
\end{equation}\end{linenomath*}
The values of initial parameters for the end-member models are summarized in Table~\ref{tab:ini_param} \add[review]{while the corresponding Mohr-Coulomb diagrams illustrating their initial stress fields can be found in the supplementary materials (see Figure S1)}. 
For a fixed angle  $\Psi$, note that because we set $S$ constant between the six models, a constant stress drop is equivalent to a constant fracture energy (equation \ref{eqn:fractureenergy}). Therefore, the characteristic length scales for the friction law, $R_0$ and $L_c$, are also constant.

%

\begin{table}
 \centering
 \begin{tabular}{|clc|cccc|}
 \hline
& \multirow{2}{*}{Parameter} & \multirow{2}{*}{Symbol} & \multicolumn{4}{c}{Initial stress field } \\ 
 \cmidrule(lr){4-7}
 & & & case (\textit{i}) &  case (\textit{vi}) &  case (\textit{vii}) &  case (\textit{xii})\\
 & & & strike-slip & reverse & strike-slip & reverse\\
 \hline
 \parbox[t]{2mm}{\multirow{4}{*}{\rotatebox[origin=c]{90}{Input}}} %
& angle &$\Psi$ & $55^o$ & $55^o$ & $15^o$ & $15^o$\\
& seismic ratio &$S$ & 2 & 2 & 1 & 1 \\
& stress drop (MPa) &$\Delta\tau$ & 10.2  & 10.2  & 10.6  & 10.6  \\
&static friction &$\mu_s$   & 0.68 & 0.52 & 0.68 & 0.52 \\
\hline
 \parbox[t]{2mm}{\multirow{9}{*}{\rotatebox[origin=c]{90}{Resulting parameters}}} %

&dynamic friction &$\mu_d$   & 0.34 & 0.16 & 0.19 & 0.09\\
& full stress tensor (MPa) &$\s{yy}$& -91 & -84 & -43 & -50\\
&&$\s{xx}$& -61 & -67 & -108 & -102\\
&&$\s{zz}$& -42 & -42 & -42& -42\\
&&$\s{xy}$& 41 & 23 & 19 & 15\\
&principal stresses (MPa)&$\s{1}$ & -120  & -101  & -113 & -106 \\
&&$\s{2}$ & -42  & -51  & -42 & -45 \\
&&$\s{3}$ & -32 & -42 & -38 & -42\\
&process zone (m) &$R_{0} $ & 1047 & 1047 & 1570 & 1570 \\
&nucleation length (m) &$L_{c} $ & 6562 & 6562 & 4374 & 4374 \\
 \hline
\end{tabular}
\caption{Initial parameters used for the two end-members of each set of models.}
\label{tab:ini_param}
\end{table}

\section{Results}
\label{sec:result}
\subsection{Role of pre-stresses in determining the yield criterion}
\label{subsec:elasmedium}

In this section,  we run simulations with an elastic medium and we compute different yield criterion commonly used in the literature to determine the inelastic deformation. It is important to note that in those cases, the medium has a pure elastic behavior and the off-fault plastic deformation is calculated a posteriori.

\subsubsection{Role of the out-of-plane stress in computing the plastic yield criterion}
\label{subsubsec:outofplane}

We first examine the importance of accounting for the out-of-plane stress in determining the off-fault plastic deformation. Here we compute the Drucker-Prager criterion:
\begin{linenomath*}\begin{equation}
F_{DP} =\sqrt{J_2} + \mu_s p
\label{eqn:DP}
\end{equation}\end{linenomath*}
where $p\equiv I_1/3=(\s{xx}+\s{yy}+\s{zz})/3$ 
is the hydrostatic stress derived from the first invariant  $I_1$ of the stress tensor and $J_2=s_{ij}s_{ij}/2$ corresponds to the second invariant of the deviatoric stress tensor (with $s_{ij} = \sigma_{ij} - p\delta_{ij}$).
However, in many of the published 2D studies \citep[][among others]{templeton2008off,dunham2011earthquake} the out-of-plane stress, $\s{zz}$, is assumed to be the mean of the in-plane stresses i.e $\s{zz} = (\s{xx}+\s{yy})/2$. This particular choice of $\s{zz}$ makes the Mohr-Coulomb and Drucker-Prager yield surfaces coincide in 2D. Therefore, the invariants can be computed using the in-plane stress tensor components i. e. $p^{ip}=(\s{xx}+\s{yy})/2$ which consequently changes the second invariant of the deviatoric stress tensor as well (further referred as $J_2^{ip}$). We note the Drucker-Prager criterion, solely using the in-plane stresses as follow:
\begin{linenomath*}\begin{equation}
F_{DP}^{ip} =\sqrt{J_2^{ip}} + \mu_s p^{ip}
\label{eqn:DPip}
\end{equation}\end{linenomath*}

\begin{figure}
\centering
\includegraphics[width=1\textwidth]{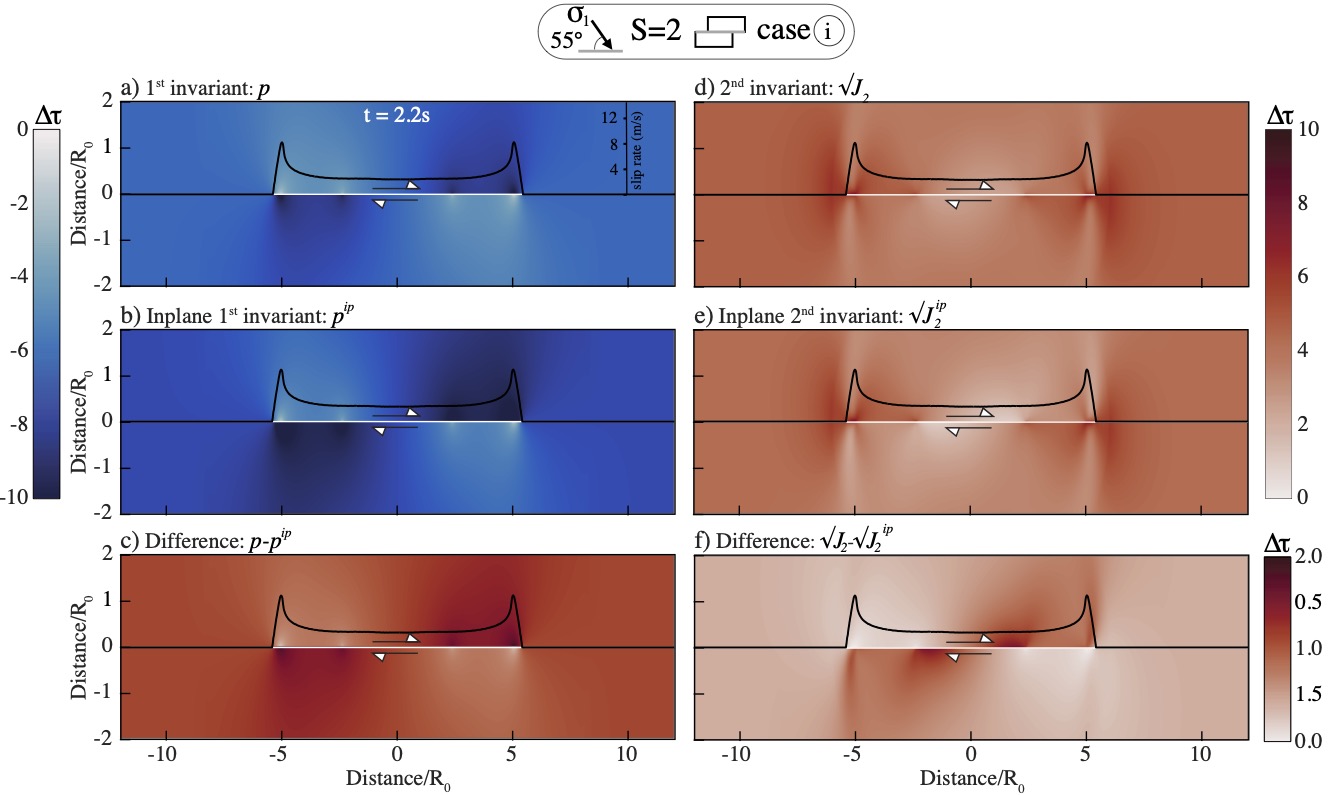}
\caption{First invariant of the stress tensor (a,b) and second invariant of the deviatoric stress tensor (d,e) computed with the 3D stress tensor (a,d), the in-plane stress tensor (b,e), at t = 2.9 seconds for case (\textit{i}) with $S = 2$ and $\Psi$ = 55 $\degree$. Figures (c,f) give respectively the difference between the way of computing the first  invariant of the stress tensor and the second invariant of the deviatoric stress tensor. Invariants are normalized by the dynamic stress drop $\Delta\tau$ (equation~\ref{eqn:stressdrop}). Slip rate on the fault (black curves) is super-imposed on the graphs.}
\label{fig04invariantsSS}
\end{figure}

\begin{figure}
\centering
\includegraphics[width=1\textwidth]{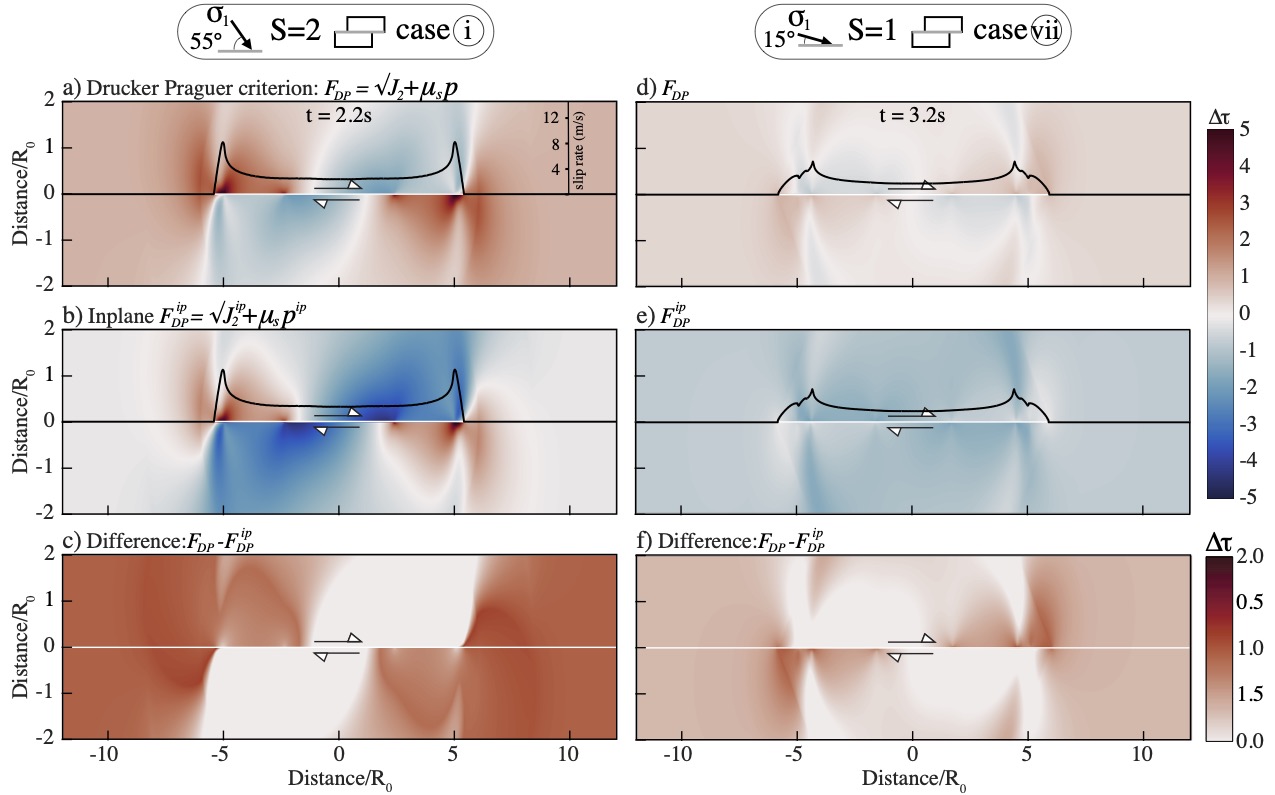}
\caption{Drucker-Prager criterion for case (\textit{i}) with $S = 2$ and $\Psi$ = 55 $\degree$ (a,b) and for case (\textit{vii}) with $S = 1$  and $\Psi = 15^o$ (d,e), at t = 2.2 seconds. We compute the invariants accounting for $\sigma_{zz}$ (a,d), or only using the in-plane stress field (b,e). We show the differences $F_{DP}- F_{DP}^{ip}$ in Figures (c,f) for cases (\textit{i}) and (\textit{vii}), respectively. The difference is computed for the positive values only. \add[review]{Different versions of the Drucker-Prager criteria and their differences} are normalized by the dynamic stress drop $\Delta\tau$ (equation~\ref{eqn:stressdrop}). Slip rate on the fault (black curves) is super-imposed on the graphs.} 
\label{fig05druckerPragerSS}
\end{figure}

To illustrate the contribution of the out-of-plane stress in computing the invariants we use case (\textit{i}) with $\Psi = 55^o$ and $S$ = 2 (see Table~\ref{tab:ini_param}).  We compare the hydrostatic stress and the square root of the second invariant of the deviatoric stress tensor obtained with and without the out-of-plane stress (Figures \ref{fig04invariantsSS}a \& b) when the fault has ruptured about 5 times the process zone $R_0$.
We can observe that in both cases, the hydrostatic stress is higher in the compressional quadrants. However, if the out of plane stress is ignored, the hydrostatic stress is overestimated (stresses are positive in tension) by up to $20$ MPa (or two \add[review]{times the} dynamic stress drop) as displayed in Figure  \ref{fig04invariantsSS}c.  On the other hand, the second invariant of the deviatoric stress tensor is underestimated if we ignore $\sigma_{zz}$ (Figures  \ref{fig04invariantsSS}d, e \& f). The difference between $J_2$ and $J_2^{ip}$ is up to $\sim$1.5 times the dynamic stress drop.
This results in a significant difference in the estimation of the Drucker-Prager criterion (Figure \ref{fig05druckerPragerSS}a, b \& c).  In both cases,  $F_{DP}$ or $F_{DP}^{ip}$ are positive in the tensional quadrants. However, taking into account the out-of-plane stress not only increases the area where the Drucker-Prager criterion is positive, i.e., where the plastic deformation is expected, \add[review]{but the overall magnitude of the plastic strain is also higher}. Hence this leads to an underestimation of the plastic deformation. 
\add[review]{As expected, Figure S2 reveals consistent findings when computing the Drucker-Prager criterion, solely using the dynamic change of stresses due to the rupture propagation.}
%

Changing the initial state of stress, using case (\textit{vii}), i.e. for $\Psi = 15^o$ and S = 1 (see Table~\ref{tab:ini_param}) we obtained even more drastic differences.  When the out-of-plane stress is ignored (Figure \ref{fig05druckerPragerSS}e) $F_{DP}^{ip}$ is pretty much negative everywhere, which may lead to the interpretation that no plastic deformation is happening.  Whereas, when accounting for $\sigma_{zz}$, even if the magnitude is about four times smaller than for  case (\textit{i}), we record positive $F_{DP}$, notably within the compressional quadrants.

\

\subsubsection{\change[review]{Role of the initial stress field for plastic yield criterion}{Role of the reverse versus strike-slip stress field on the plastic yield criterion}}
\label{subsubsec:plasticcriterion}

Now that we have emphasized the importance of including the full stress tensor in computing the Drucker-Prager criterion, we look at the influence of setting up a proper strike-slip initial stress field, ensuring that the pre-stress conditions satisfy $\s{1}<\s{2}=\s{zz}<\s{3}$. 

\begin{figure}
\centering
\includegraphics[width=1\textwidth]{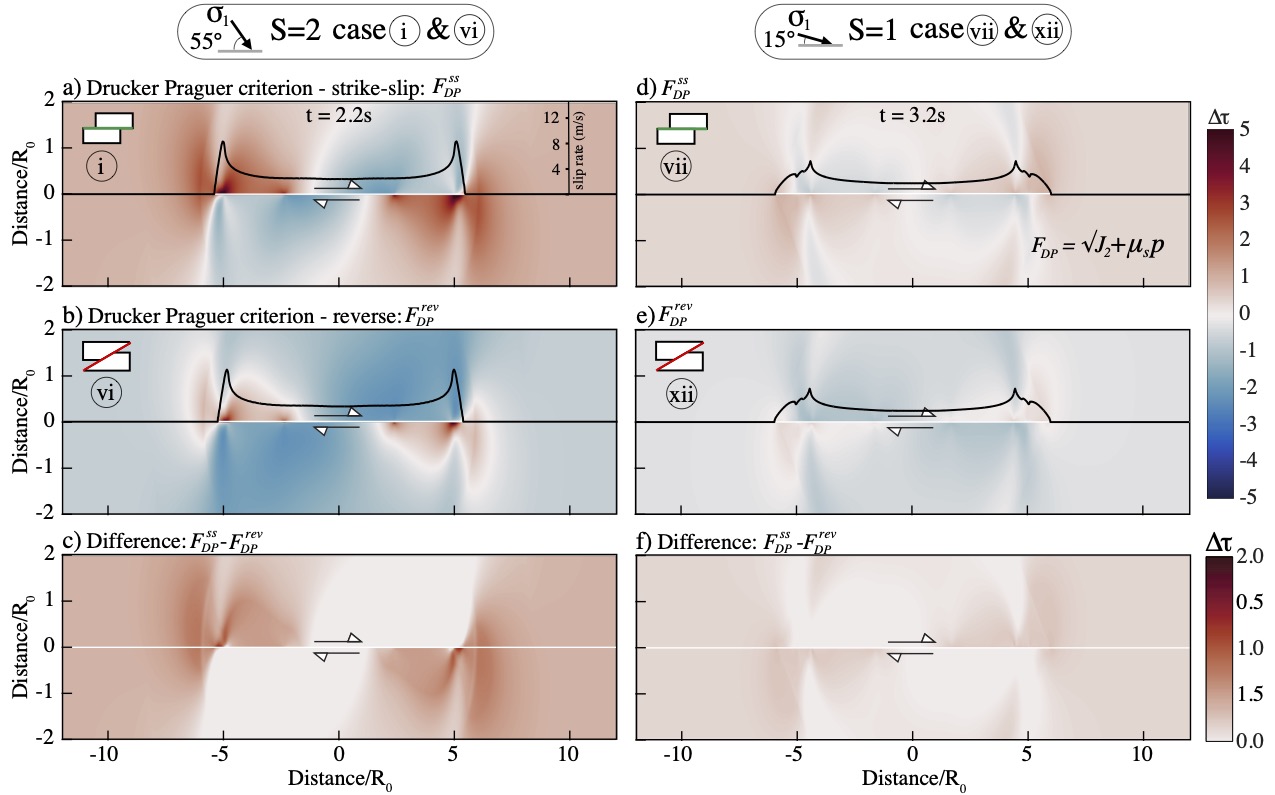}
\caption{Drucker-Prager criterion for cases (\textit{i}) \& (\textit{vi}), with $S = 2$ and $\Psi$ = 55 $\degree$ (a,b) and for cases (\textit{vii})  \& (\textit{xii}), with $S = 1$  and $\Psi = 15^o$ (d,e). 
Cases (\textit{i}) \& (\textit{vii}) correspond \remove[review]{s} to an initial strike-slip stress field (a,d) and cases (\textit{vi})  \& (\textit{xii}) to an initial reverse stress field (b,e).  Figures (c,f) give the difference between cases (\textit{i}) \& (\textit{vi}), and between cases (\textit{vii})  \& (\textit{xii}), respectively. The four models are plotted against the criterion for accurate initial stress in Figure \ref{fig03condition}. Drucker-Prager criterions and differences are normalized by the dynamic stress drop $\Delta\tau$ (equation~\ref{eqn:stressdrop}). Slip rate on the fault (black curves) is super-imposed on the graphs.} 
\label{fig06DP_ss-rev}
\end{figure}

In Figure \ref{fig06DP_ss-rev},  for $\Psi = 55^o$, $S$ = 2 and for $\Psi = 15^o$, $S$ = 1  we compare the  Drucker-Prager criterion of the end-member models, i.e  case (\textit{i}), with case (\textit{vi}) and  case (\textit{vii}) with case (\textit{xii}).  The pre-stress conditions for cases (\textit{i}) and (\textit{vii}) correspond to a strike-slip fault,  whereas cases (\textit{vi}) and (\textit{xii})  have \remove[review]{a} the initial stress field of a reverse fault.
The yield criterion is computed when the fault has ruptured about 5 times the process zone $R_0$.  \add[review]{Note that, due to the difference in S ratio for the simulations with $\Psi = 55^o$ and $\Psi = 15^o$, it is not meaningful to compare the models with different angles, as the dynamic rupture is very different, and by extent, the off-fault deformation.} 
 
For initial reverse stress conditions, when $\Psi = 55^o$ (Figure \ref{fig06DP_ss-rev}b, case \textit{vi}), positive $F_{DP}$ is observed in the tensile quadrants.  When $\Psi = 15^o$ (Figure \ref{fig06DP_ss-rev}e, case \textit{xii}), positive $F_{DP}$ is observed in the compressive quadrants, with a magnitude lower than for case (\textit{vi}). 
When the initial stress field is properly set-up (Figure \ref{fig06DP_ss-rev}a \& d), the areas recording a positive $F_{DP}$ are much larger and the absolute value is also higher,  as illustrated by Figure \ref{fig06DP_ss-rev}c \& f. 
\add[review]{Since the models are run within a purely elastic medium and since the rupture dynamics are very similar,  the differences are essentially linked to the background stresses}.

\begin{figure}
\centering
\includegraphics[width=1\textwidth]{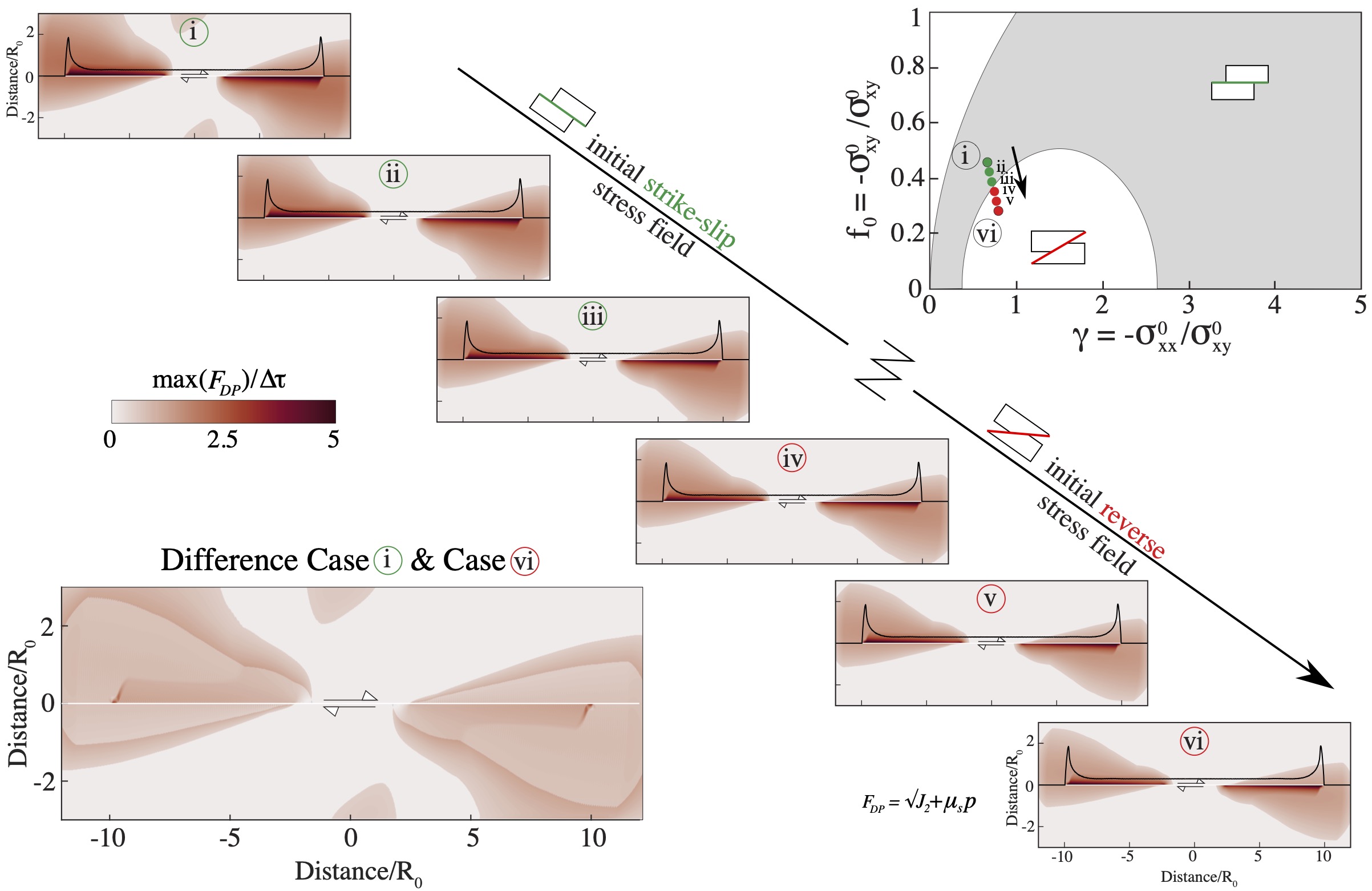}
\caption{Drucker-Prager criterion for all models with $\Psi = 55^o$ and S=2. We plot the maximum value $F_{DP}$  between t=0 and t=4 seconds. Only positive values \add[review]{ are considered (hence the color discontinuity.) } . We display the results from the strike-slip end-member (upper left corner) to the reverse end-member (lower right corner). The initial parameters are shown in the upper right inset, in the same representation \change[review]{than}{as in} Figure \ref{fig03condition}a. The lower left inset shows the difference between the two end members.  Drucker-Prager criterions and differences are normalized by the dynamic stress drop $\Delta\tau$ (equation~\ref{eqn:stressdrop}). Slip rate on the fault (black curves) is super-imposed on the graphs.} 
\label{fig07DPdiffPhi55S2}
\end{figure}

Figure \ref{fig07DPdiffPhi55S2} shows the continuum of all models for $\Psi= 55^o$.  Unlike Figure \ref{fig06DP_ss-rev} for which we plot$F_{DP}$ at one particular time step, here we plot the maximum value of $F_{DP}$ induced  in the off-fault medium by the full rupture. 
Crossing the boundary between initial strike-slip stress field and reverse pre-stress conditions does not change the results qualitatively. However, the area with positive $F_{DP}$ is larger in the strike-slip case, and we observe higher absolute values, with up to twice the dynamic stress drop.

\subsubsection{Role of the initial stress field for off-fault rupture modes}
\label{subsubsec:CFF}

\begin{figure}
\centering
\includegraphics[width=1\textwidth]{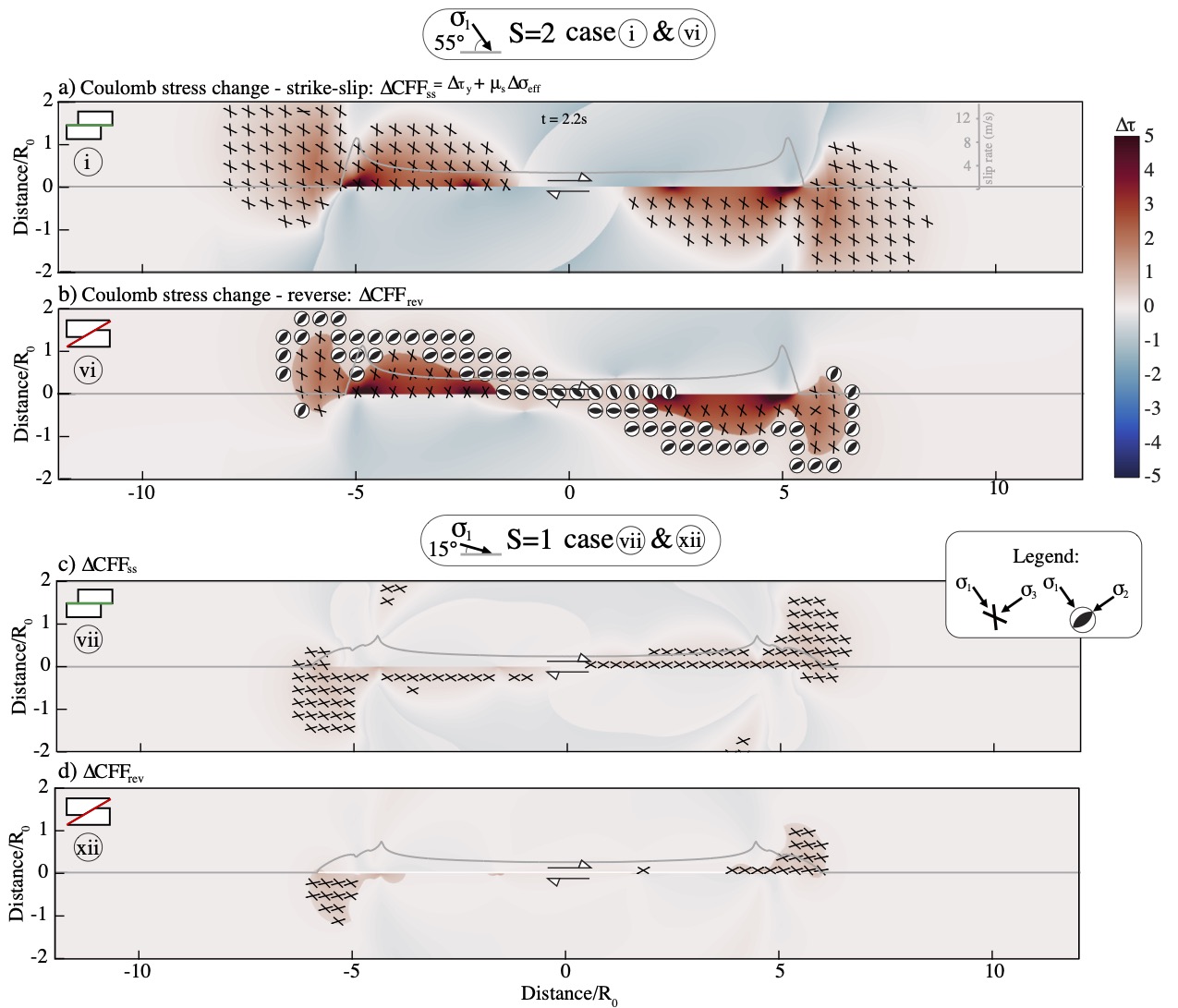}
\caption{Coulomb stress change $\Delta CFF$ due to the dynamic rupture, computed on optimally oriented planes and normalized by the dynamic stress drop $ \Delta\tau$ (equation~\ref{eqn:stressdrop}), for cases (\textit{i}) \& (\textit{vi}), with $S = 2$ and $\Psi$ = 55 $\degree$ (a,b) and for cases (\textit{vii})  \& (\textit{xii}), with $S = 1$  and $\Psi = 15^o$ (c,d). Cases (\textit{i}) \& (\textit{vii}) correspond to an initial strike-slip stress field (a,c) and cases (\textit{vi})  \& (\textit{xii}) to an initial reverse stress field (b,d). Superimposed on this snapshot are the conjugates planes that give a maximum value of $\Delta CFF$. The symbols inform on the expected mode of rupture: two crossed lines represent the two conjugate planes for strike-slip ruptures, the beach ball with a dark center gives the orientation of the conjugate planes for reverse faulting, following the classic seismological convention. Those symbols are displayed in the areas where $\Delta CFF >0.2$.  Slip rate on the fault (grey curve) is super-imposed on the graphs.} 
\label{fig08CFF}
\end{figure}

We have shown that \change[review]{pre-}{the initial} stress conditions influence the values of $F_{DP}$ criterion induced by the rupture.  Yet the Durcker-Prager criterion uses the invariants of the stress tensor, which does not predict the preferential orientation for failure,  a useful information in seismic risk assessment for example. Another criterion often used in the literature is therefore the Coulomb stress change due to the rupture on the main fault
\citep{stein1997progressive,king1994coulomb,thomas2017rate,canitano2021inherited}.
\begin{linenomath*}\begin{equation}
    \Delta CFF = (\Delta\tau_y + \mu_s \Delta\s{eff})
\label{eqn:CFF}
\end{equation}\end{linenomath*}
where $\Delta\s{t}$ and $\Delta\s{n}$ are  the change of shear and normal stress respectively \add[review]{on the optimally oriented plane}, induced by the seismic rupture for a given direction $\theta$ with respect to the principal stress $\s{1}$ (in relation to the Mohr-Coulomb circle displayed in Figure~\ref{fig01orientation}).
In Figure~\ref{fig08CFF} we explore the same cases that in Figure~\ref{fig06DP_ss-rev}, i.e.,  the two end-members for $\Psi = 55^o$, S = 2 (cases \textit{i} \& \textit{vi}) and for $\Psi = 15^o$, S = 1 (cases \textit{vii} \& \textit{xii}).  
The background color corresponds to local values of $\Delta CFF$.
The yield criterion is computed when the fault has ruptured about 5 times the process zone $R_0$.
As observed previously with the Drucker-Prager criterion, areas with positive \change[review]{c}{C}oulomb stress change are larger and with higher $\Delta CFF$ values for $\Psi = 55^o$ \add[review]{Figures}~\ref{fig08CFF}a \& b.  Likewise,  the areas likely to induce off-fault deformation are larger when the initial stress-field is properly set up, and this for the two tested values of $\Psi$  \add[review]{Figures}~\ref{fig08CFF}a \& c. 

On top of these snapshots,  we compute the local preferential orientations for failure and the corresponding type of faulting induced.
An off-fault strike-slip failure means that locally,  $\s{1}$ and $\s{3}$ are in-plane. If $\s{3}$ is out-of-plane, the local preferred type of failure will be that of a reverse fault. 
Figure \ref{fig08CFF}a shows that if the initial strike-slip stress field is correctly set up, the rupture induces only strike-slip off-fault failures. However, if the initial stress field actually corresponds to that of a reverse faulting, the outcomes \remove[review]{is} are different (Figure \ref{fig08CFF}\remove[review]{c}b).  The main rupture (strike-slip by default since we are in 2D) only influences the stress field close to the rupture tips.  Far from the fault and within the nucleation area, the propagating rupture does not control the local stress field,  but its initial value does. Hence we observe off-fault reverse failures for case (\textit{vi}).


\subsection{Dynamic simulation with inelastic rheology}
\label{subsec:inelastic}

We have shown, using \change[review]{pure}{linear} elasticity, that the pre-stress conditions can significantly affect the assessment of the different yield criteria used to estimate the off-fault deformation. In \remove[review]{the following} this section,  we now investigate the role of the initial stress field on the dynamically triggered off-fault deformation, and \remove[review]{the counter-impact} its feedback on the seismic rupture, using different modelling strategies.


\subsubsection{Role of the initial stress field \change[review]{for}{in} plastic deformation}
\label{subsubsec:plastic}
We first use the off-fault plasticity model implemented in SEM2DPack following \citet{andrews2005rupture}.
The inelastic response of the medium is characterised by a Coulomb criterion using in-plane stresses with $\mu = 0.75$ and $c$ = 30 MPa so that the initial stress state of the medium is below the yield criterion.
\add[review]{The maximum shear stress over all orientations is :}

\begin{linenomath*}\begin{equation}
   \tau = \sqrt{ \sigma^2_{xy} + \left[ (\sigma_{xx} - \sigma_{yy})/2 \right]^2}
\label{eqn:plasticAndrews}
\end{equation}\end{linenomath*}

\add[review]{The Coulomb criterion is :}

\begin{linenomath*}\begin{equation}
   \tau \leq c~\cos{\mu} - 0.5(\sigma_{xx} + \sigma_{yy})\cos{\mu}
\label{eqn:coulomb_criterion}
\end{equation}\end{linenomath*}

At each time step of the calculation, stress components are first incremented elastically. Then, if the Coulomb criterion is violated, stress components are recomputed so that a part of the deformation is accommodated inelastically (see \citet{andrews2005rupture,duan2008inelastic} for details about the method).  

\begin{figure}
\centering
\includegraphics[width=0.7\textwidth]{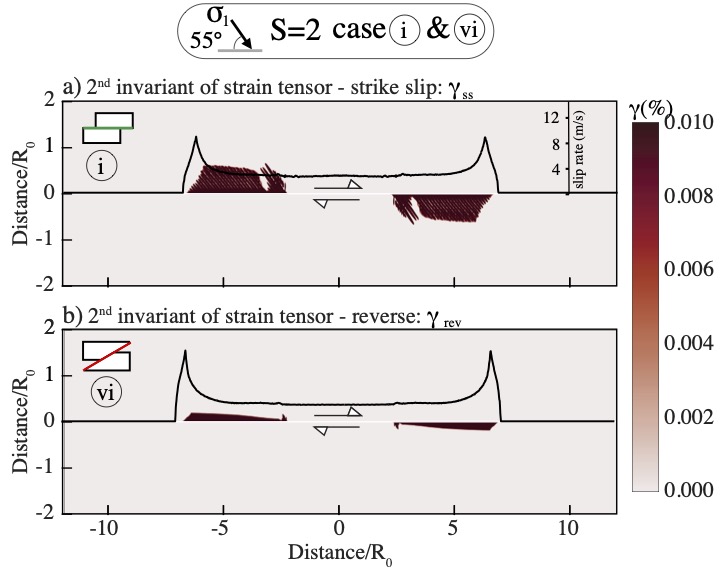}
\caption{Plastic deformation for the models with $\Psi = 55^o$, S=2, with initial strike-slip stress field for case (\textit{i}) and reverse stress field for case (\textit{vi}) . Here we show the second invariant of the deviatoric strain tensor $\gamma$. Slip rate on the fault (black curve) is super-imposed on the graphs.} 
\label{fig09plasticity}
\end{figure}
\begin{figure}
\centering
\includegraphics[width=1\textwidth]{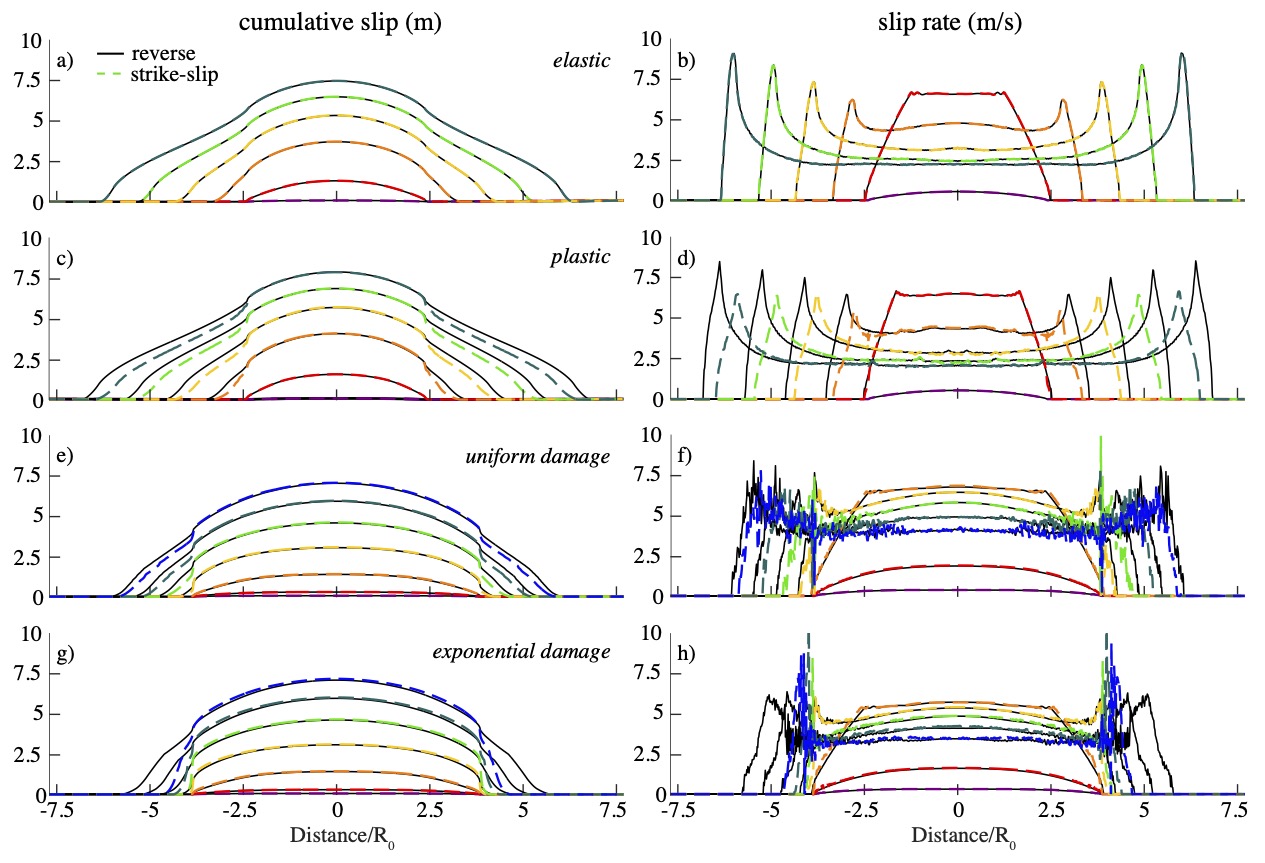}
\caption{Cumulative slip (a-c-e-g) and slip rate (b-d-f-h) plotted every 0.4 seconds for $\Psi = 55^o$ and $S=2$, with initial reverse stress fields (dark solid lines,  case \textit{vi}) and initial strike-slip stress field (dotted color lines, case \textit{i}). } 
\label{fig10slips}
\end{figure}

%
We compute the plastic deformation induced by the rupture for the two end-members cases (\textit{i}) and (\textit{vi}) previously studied ($\Psi = 55^o$ and $S$=2).  In Figure~(\ref{fig09plasticity}),the results are illustrated by the second invariant of the deviatoric strain tensor $\gamma_{t}$:
\begin{linenomath*}\begin{equation}
    \gamma_{t} = \sqrt{2e_{ij}e_{ij}}
\label{eqn:gamma}
\end{equation}\end{linenomath*}
where $e_{ij} = \epsilon_{ij} - \frac{1}{3} \epsilon_{kk} \delta_{ij}$.
The pattern of off-fault deformation is significantly different between the two simulations. The extent of the plastic deformation is much larger for an initial strike-slip stress state (Figure~\ref{fig09plasticity}a), as expected based on the results of section~\ref{subsec:elasmedium}.
This induces differences in the rupture dynamics with a decrease of rupture speed and slip rate (Figure \ref{fig10slips}d) and a lower cumulative slip (Figure \ref{fig10slips}c), compared to case (\textit{vi}) that has an initial stress field that corresponds to reverse faulting. Therefore, when off-fault inelastic deformation is taken into account, even \change[review]{small}{modest} differences in the initial stress condition affect significantly both the evolution of the off-fault medium and the slip dynamics.

We note that in a non-cohesive medium ($c$=0), those differences are even more emphasised (Figure S3a \& b). For initial strike-slip stress conditions, the rupture decelerates rapidly, its propagation is prevented by the intense inelastic deformation of the medium. The off-fault deformation is localised and optimally oriented with respect to the far field stress orientation.  It is equivalent to the creation of a new optimally oriented fault.  Hence, it is interesting to note that modelling off-fault deformation for a 2-D strike-slip fault, with the appropriate initial stress field,  requires a certain cohesion in order to fully rupture the prescribed fault plane.
We also note that for a non-cohesive medium, when $\Psi= 15^o$ and $S$=1 (Figure S3c \& d)., little deformation occurs for the strike-slip case (\textit{vii}) and none for the reverse case (\textit{xii}). When $\Psi = 30^o$ (optimally oriented fault, Figure S3e \& f),  deformation only occurs on the main fault. Therefore,  a flat fault has to be mis-oriented to produce significant off-fault deformation.

\subsubsection{Role of the initial stress field on dynamic damage}
\label{subsubsec:damage}

The last set of simulations use a micromechanics-based model to determine the dynamically-triggered off-fault damage and its \change[review]{counter-impact}{feedback} on the rupture dynamic. Inelastic deformation can occur in the model by either crack opening or crack propagation from initial flaws. Using an energy-based approach, at each time step, the corresponding change in elastic moduli, and hence the constitutive law, is determined (see \cite{thomas2017effect} \add[review]{for constitutive equations}).  The current inelastic state of the medium is \change[review]{defined}{characterized} by the scalar $D$, the fraction of volume occupied by microcracks:
\begin{linenomath*}\begin{equation}
    D=\dfrac{4\pi}{3}N_v\left(a\cos\Phi_c+l\right)^3
\label{eqn:D}
\end{equation}\end{linenomath*}
where $a$ is the initial microcrack radius,  $l$, the wings cracks length as they are growing parallel to $\si{1}$  ($l=0$ at $t=0$) (see figure 11) and $N_v$ the volume density of cracks. Initial flaws are all aligned at the same optimal angle  $\Phi_c=\frac{1}{2}\tan^{-1}(1/\mu_c)$ to $\si{1}$, with $\mu_c$ being the friction coefficient for the microcracks. \add[review]{In their model,} \citet{bhat2012micromechanics} \add[review]{derive the crack growth law by comparing the stress intensity factor at the  microcrack tips to the experimentally-determined initiation and propagation toughness.}

$D$ varies between 0 and 1, the maximum value corresponding to the coalescence stage that leads to the macroscopic fracture of the solid. See \citet{bhat2012micromechanics} and \cite{thomas2017effect} for further details on the method. 

\begin{figure}
\centering
\includegraphics[width=1\textwidth]{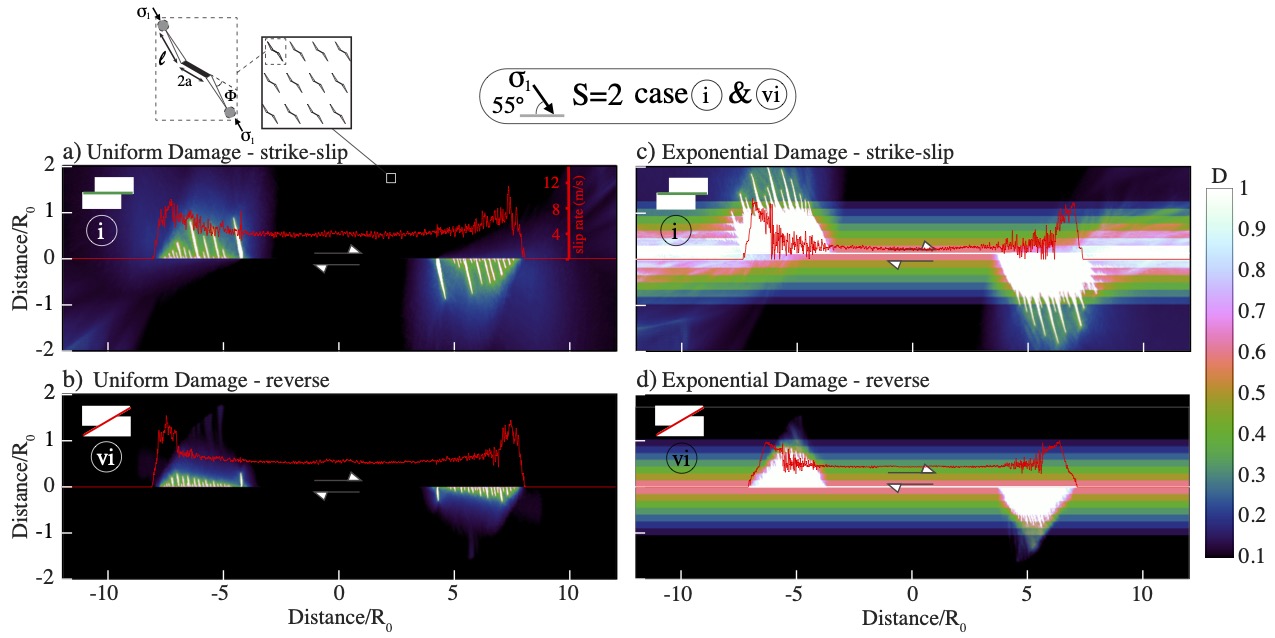}
\caption{Simulation of a dynamic rupture with off-fault damage within an homogenous medium (a,b) and within a medium a pre-existing damage zone, with an exponential decay of damage density away from the fault (c,d). We explore the two end-member cases for  $S = 2$ and $\Psi$ = 55 $\degree$: case (\textit{i}) corresponds to an initial strike-slip stress field (a,c) and  cases (\textit{vi}) to an initial reverse stress field (b,d). The colors represent the density of microcracks in the medium. Slip rate on the fault (red curve) is super-imposed on the graphs. The inset shows a schematic representation of the initial crack distribution and of the crack geometry (after \citet{thomas2017effect}).}
\label{fig11damage}
\end{figure}

Figure \ref{fig11damage} shows the damage density induced by the rupture for the end-member cases (\textit{i}) and (\textit{vi}) with $\Psi = 55^o$ and S=2.  The distribution of pre-existing flaws is homogeneous ($D=0.1$ at $t=0$) in Figures  \ref{fig11damage}a \& b.
For the models in Figures \ref{fig11damage}c \& d we assume an exponential decay of initial damage with fault normal distance, as described in several field studies \citep[e.g.,][]{Vermilye1998,Wilson2003,mitchell2009nature}. The initial damage density varies from $D=0.5$ to $D=0.1$ over a distance equivalent to the process zone $R_0 \sim 1$ km. \add[review]{The characteristic scale was selected to match the width of the damage zone that may have been created by past earthquakes, in this case defined in }Figure \ref{fig11damage}a.
In all scenarios, to prevent off-fault damage at the beginning of the simulation due to far field loading, we set the friction on the microcracks, $\mu_c$ to 0.75.  Hence the observed damage is dynamically triggered by the seismic rupture.

Similar to the models discussed earlier, for these particular stress states,  damage essentially occurs in the tensional quadrants.  The damage zone is also significantly wider for initial strike-slip stress conditions (Figures \ref{fig11damage}a \& c).
Previous studies, in comparison to simulations with a pure elastic medium,  have underlined the effect of damage on slip rate and rupture velocity (slow down) and to a lesser extent the cumulative slip \citep{thomas2017effect,thomas2018dynamic}.
Here, when comparing in Figure \ref{fig10slips} the models with initial reverse stress fields (dark solid lines) and strike-slip stress field (dotted color lines), we observe differences in cumulative slip (e \& g), slip rate and rupture velocity (f \& h), both for models with an uniform medium or an initial damage zone. That is because, unlike for the reverse case, when the initial stress field is correctly set up, damage occurs ahead of the rupture tip, thus changing the P- and S-wave speeds in the medium, which ultimately slows down the rupture velocity. This effect is even more pronounced when the earthquake ruptures a fault with a pre-existing damage zone (Figures \ref{fig10slips}g \& h).

\section{\change[review]{Conclusion}{Discussion and Conclusions}}
\label{sec:discussion}


\begin{figure}
\centering
\includegraphics[width=1\textwidth]{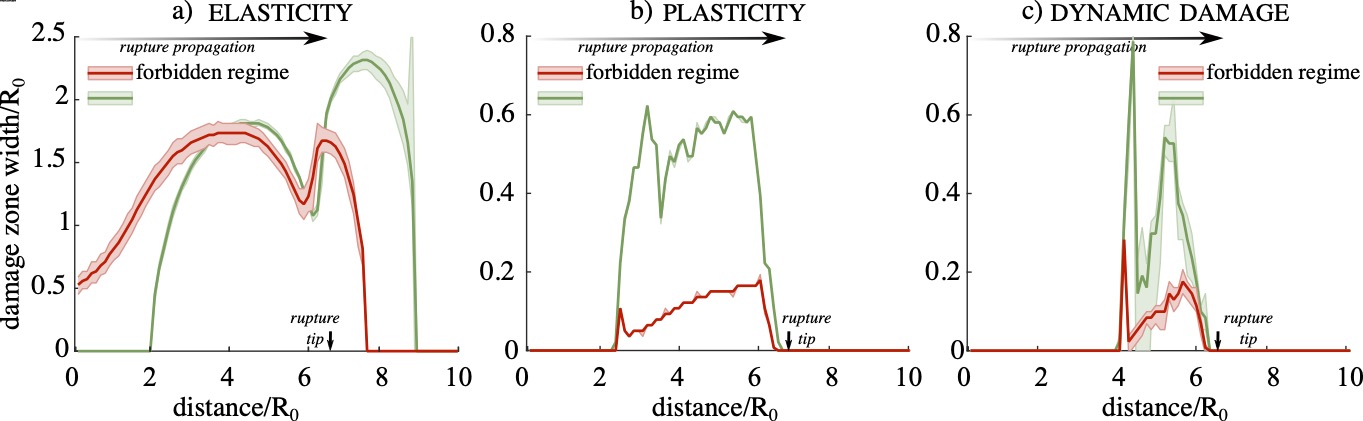}
\caption{Synthesis of the effect of initial stress on damage zone width for cases (\textit{i}) and (\textit{vi}) with $\Psi = 55^o$ and S=2, when the rupture has reached seven times the \add[review]{static} process zone size. Results from the different models are displayed in red for case (\textit{vi}), with reverse pre-stress conditions, and in green for case (\textit{i}), with strike-slip pre-stress conditions.  a) Expected damage zone width using purely elastic models. We use the Coulomb stress change with a threshold of $\Delta CFF = 0.3 \Delta\tau$. b) Damage zone width computed with the model following \citet{andrews2005rupture}, using a threshold value of plastic deformation of $0.01 \%$. c) Damage zone width computed with the micromechanical model, using a threshold value of $D=0.3$ (medium with a uniform initial damage density $D_0=0.1$). The shaded areas correspond to $\pm 10 \%$ changes in the threshold value.}
\label{fig12synthese}
\end{figure}

As discussed above,  in the last two decades, the wealth of observations have underlined the importance of off-fault deformation and complex structure in fault zone behavior.  Numerical models have been developed to incorporate these key ingredients \citep[][among others]{andrews2005rupture,Shi2006,templeton2008off,dunham2011earthquake,thomas2017effect,okubo2019dynamics}. However, due to numerical limitations, they have been developed mostly in 2D. As a consequence, when setting up the initial stress field, only the in-plane stresses are defined and the out of plane stress is often ignored, or assumed to be the mean of the in-plane stresses.
%
%

In several studies, the effect of initial stress state have been investigated in terms of  the orientation of $\sigma_1$ with respect to the fault direction. They illustrated the significant influence of $\Psi$ on the pattern of off-fault damage \citep{poliakov2002dynamic,rice2005off,ngo2012off,templeton2008off}. However, the impact of the relative importance of the principal stresses, at constant angle $\Psi$, has not been discussed. In this study, we run 2D plane-strain simulation of a strike slip faulting to illustrate the key role of the 3D faulting regime on off-fault deformation.

Using an elastic medium, we first show that, even if the initial stress field is rightfully set up, ignoring the out-of-plane stress (here $\s{zz}$)  leads to an underestimation of the inelastic deformation, both in extend and magnitude.  If the initial stress field is on top wrongly defined (reverse faulting), ignoring $\s{zz}$ will on contrary lead to an over-estimation of the inelastic deformation. 
Then, we demonstrate that a \change[review]{small}{modest} change in the pre-stress conditions, from strike-slip to reverse stress field, strongly influences the magnitude of any plastic criterion and the extent of off-fault deformation.  Using a Coulomb criterion, we also compute the local preferential orientations for failure and the corresponding type of faulting induced. We have shown that a simulation within the so-called ``forbidden regime" will predict some reverse faulting in the off-fault medium. 

Then,  because of the various feedbacks that exist between the dynamic rupture and the bulk, we show that the discrepancy is even more pronounced when inelastic deformation can occur in the medium as the rupture propagates (figure~\ref{fig12synthese}). Both in plastic and dynamic damage models, the resulting pattern of inelastic deformation is significantly different. We show that a modest change in pre-stress conditions from initial reverse to initial strike-slip stress field would underestimate the damage zone width by a factor of 3 to 6 (Figure \ref{fig12synthese}b,c). 
We would like to underline that, while previous numerical studies have modelled inelastic deformation under different stresses regimes \citep{Shi2006,dunham2011earthquake,templeton2008off,okubo2019dynamics},  as illustrated in Figure~\ref{fig03condition}a, they did not investigate the effect of pre-stress by keeping S ratio, stress drop, and angle $\Psi$ simultaneously constant. By keeping these rupture parameters constant among our models, we demonstrate the importance of initial stress field only. 
We also show that the effect of far field stresses can have a significant impact on the rupture dynamics (Figure~\ref{fig10slips}). In a passive, elastic medium, pre-stress indeed does not affect fault slip. In inelastic medium, such as displayed in Figures \ref{fig09plasticity} and \ref{fig11damage} the evolving medium through energy loss and/or trapped-waves influences back both the slip rate and rupture velocity on the fault. We observe that this effect is more predominant with higher amount of off-fault deformation, hence when the initial stress field is rightfully set up and/or if a pre-existing damage zone is modelled. Therefore in 2D plane-strain simulations, the initial three dimensional stress field is important to model an accurate evolution of the off-fault medium.

To conclude, pre-stresses can significantly affect both off-fault damage and on-fault rupture dynamics even if other key parameters are kept constant: cohesive zone, nucleation size, seismic ratio, stress drop.
Although none of the presented numerical models are meant to reproduce field observations exactly the sheer increase in observations opens up the potential for statistical comparisons between models and observations.  This makes it even more urgent to set-up the correct initial ``3D'' stress field even in ``2D'' numerical simulations.

\section*{Acknowledgments}
This study was supported by the Agence National de la Recherche (ANR) IDEAS contract ANR-19- CE31-0004-01. H. S. B. acknowledges the European Research Council grant PERSISMO (grant 865411) for partial support of this work. We thank Prof. Shiqing Xi and an anonymous reviewer for constructive feedback that greatly helped us in conveying our message. We would also like to thank our editor Prof. Fukuyama for his helpful insights in the process. 
\section*{Data Availability}
This study uses numerical data only. All of the models and data sets are produced by the authors. The numerical model used to perform the simulations \citep{ampuero2012sem2dpack} is available at the following link {\url{https://github.com/jpampuero/sem2dpack}}. The module developed by \cite{thomas2017effect} to simulate dynamic damage will be shared on request to the corresponding author. We provide a Matlab code to check a given initial stress field and plot Figure 3 {\url{https://github.com/louisejeandetribes/PreStress_2Dplanestrain.git}}


\clearpage
\renewcommand\thefigure{S\arabic{figure}}    
\setcounter{figure}{0}  
\renewcommand\thesection{S\arabic{section}}    
\setcounter{section}{0} 
\clearpage  

\begin{center}
{\Large {\textbf{\centering Supplementary Figures}}}
\end{center}

\section{Initial stress field}
This section provides the Mohr-Coulomb diagrams that illustrate the initial stress fields of the four main models of the manuscript.

\section{Elastic models - dynamic stress changes}
This section displays the Drucker-Prager criterion for the model with $\Psi = 55^o$ and $S=2$, computed with the whole stress tensor and the in-plane stresses only.


\section{Supplementary models with plastic deformation}
This section provides the supplementary figure cited in section \textit{4.2.1 : Role of the initial stress field on plastic deformation}.

\clearpage

\begin{figure}
\centering
\includegraphics[scale=0.4]{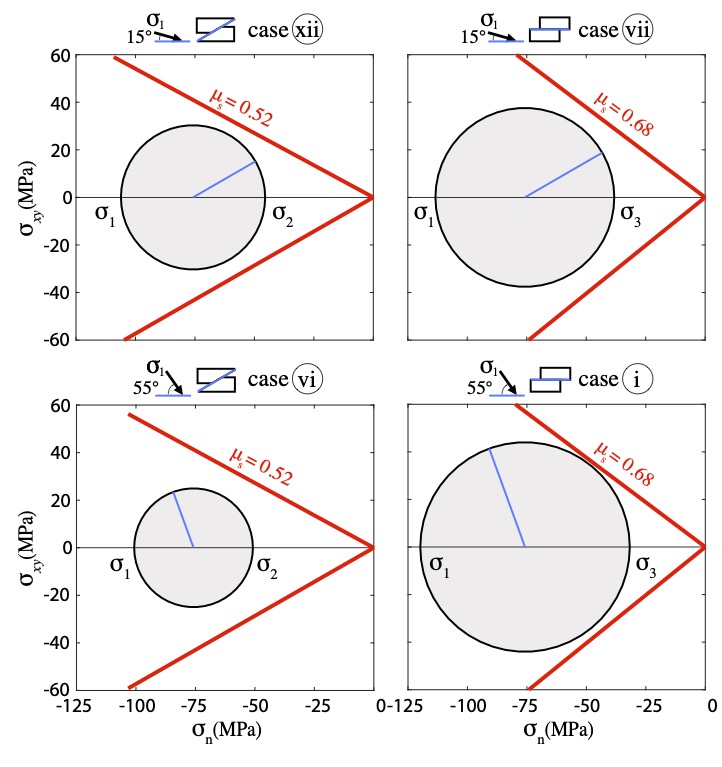}
\caption{Initial stress fields of the four main models. The Mohr Circle is depicted between $\sigma_1$ and $\sigma_2$ when the initial stress field is incorrect (i.e.,  it actually correspond to a reverse stress field in 3D), and between $\sigma_1$ and $\sigma_3$ when the initial stress field is correct (i.e., it  does correspond to a strike-slip stress field in 3D). It is important to notice that we artificially trigger the rupture by applying a higher shear stress on a nucleation patch that is not included in this figure.} 
\end{figure}

\begin{figure}
\centering
\includegraphics[scale=0.3]{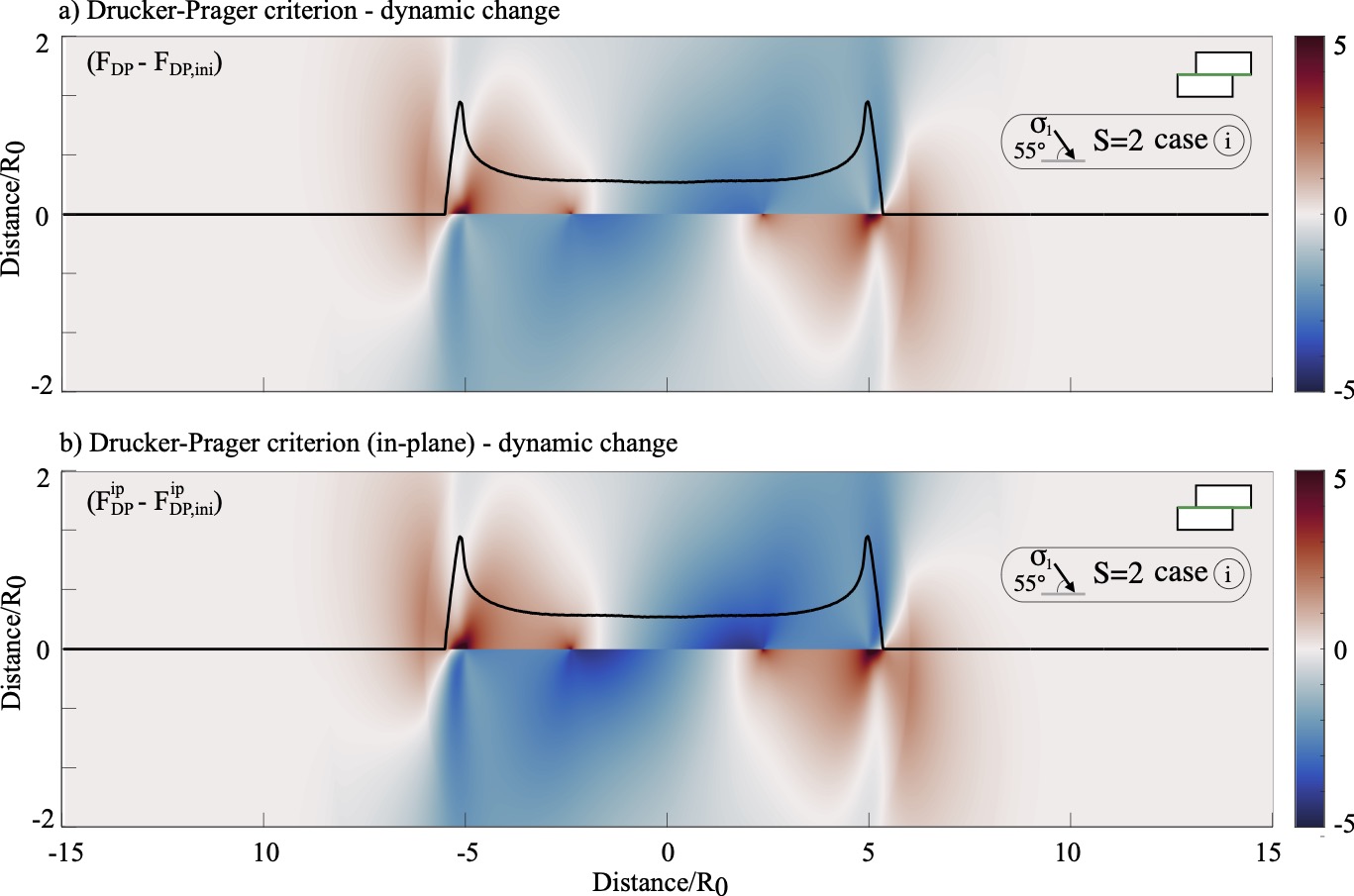}
\caption{Drucker-Prager criterion for the model with $\Psi = 55^o$ and $S=2$, computed with the whole stress tensor (a) and the in-plane stresses only (b).  Contrary to Figure 5,  here we only show the 
dynamic change of plastic criterion by removing the contribution from the initial stress field $(F_{DP}-F_{DP,ini})$.  Be careful, here the simulations are identical.  Only the error in computing the Drucker-Prager criteria, by solely using the in-plane stresses for case (b),  is responsible for the discrepancy.}
\end{figure}

\begin{figure}
\centering
\includegraphics[scale=0.3]{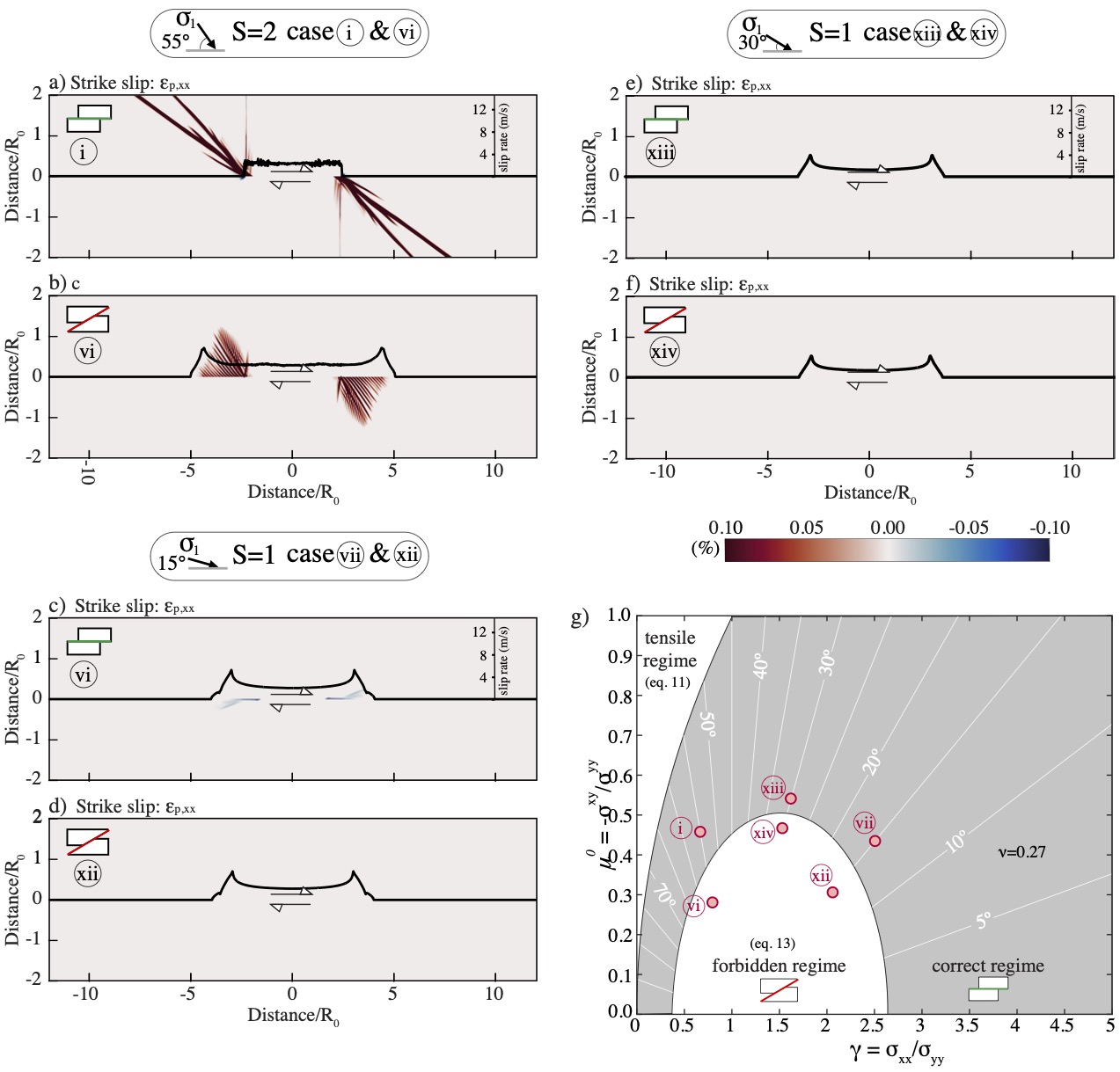}
\caption{(a) \& (b),  plastic deformation  for the models with $\Psi = 55^o$, S=2, with initial strike-slip stress field (case \textit{i}) and reverse stress field (case \textit{iv}).  (c) \& (d), plastic deformation  for the models with $\Psi = 15^o$, S=1, with initial strike-slip stress field (case \textit{vi}) and reverse stress field (case \textit{xii}).  (e) \& (f), plastic deformation for the models with $\Psi = 30^o$, S=1, with initial strike-slip stress field (case \textit{xiii}) and reverse stress field (case \textit{xiv}).  In all cases the cohesion $c = 0$.  (g) Criterion for accurate initial stress field under plane-strain approximation displayed as $\gamma-\mu_0$.  Red circles gives the initial parameters for the four simulations.} 
\end{figure}


\begin{thebibliography}{106}
\expandafter\ifx\csname natexlab\endcsname\relax\def\natexlab#1{#1}\fi

\bibitem[Abdelmeguid et~al.(2019)Abdelmeguid, Ma, \& Elbanna]{Abdelmeguid2019}
Abdelmeguid, M., Ma, X., \& Elbanna, A., 2019.
\newblock A novel hybrid finite element-spectral boundary integral scheme for
  modeling earthquake cycles: Application to rate and state faults with
  low-velocity zones, {\it Journal of Geophysical Research: Solid Earth\/},
  {\bf 124}(12), 12854--12881.

\bibitem[Ampuero(2012)]{ampuero2012sem2dpack}
Ampuero, J., 2012.
\newblock Sem2dpack, a spectral element software for 2d seismic wave
  propagation and earthquake source dynamics, v2. 3.8.

\bibitem[Anderson(1951)]{anderson1951dynamics}
Anderson, E., 1951.
\newblock The dynamics of faulting and dyke formation with applications to
  britain: Edinburgh, {\it Oliver and Boyd\/}, {\bf 206}.

\bibitem[Anderson(1905)]{anderson1905dynamics}
Anderson, E.~M., 1905.
\newblock The dynamics of faulting, {\it Transactions of the Edinburgh
  Geological Society\/}, {\bf 8}(3), 387--402.

\bibitem[Andrews(2005)]{andrews2005rupture}
Andrews, D., 2005.
\newblock Rupture dynamics with energy loss outside the slip zone, {\it Journal
  of Geophysical Research: Solid Earth\/}, {\bf 110}(B1).

\bibitem[Andrews(1976)]{andrews1976}
Andrews, D.~J., 1976.
\newblock Rupture velocity of plane strain shear cracks, {\it J. Geophys.
  Res.\/}, {\bf 81}(B32), 5679--5689.

\bibitem[Barbot et~al.(2012)Barbot, Lapusta, \& Avouac]{Barbot2012}
Barbot, S., Lapusta, N., \& Avouac, J.-P., 2012.
\newblock Under the hood of the earthquake machine: Toward predictive modeling
  of the seismic cycle, {\it Science\/}, {\bf 336}(6082), 707--710.

\bibitem[Ben-Zion \& Rice(1997)]{Ben-Zion1997}
Ben-Zion, Y. \& Rice, J.~R., 1997.
\newblock Dynamic simulations of slip on a smooth fault in an elastic solid,
  {\it Journal of Geophysical Research: Solid Earth\/}, {\bf 102}(B8),
  17771--17784.

\bibitem[Ben-Zion \& Shi(2005)]{ben2005dynamic}
Ben-Zion, Y. \& Shi, Z., 2005.
\newblock Dynamic rupture on a material interface with spontaneous generation
  of plastic strain in the bulk, {\it Earth and Planetary Science Letters\/},
  {\bf 236}(1-2), 486--496.

\bibitem[Ben-Zion et~al.(2003)Ben-Zion, Peng, Okaya, Seeber, Armbruster, Ozer,
  Michael, Baris, \& Aktar]{ben2003shallow}
Ben-Zion, Y., Peng, Z., Okaya, D., Seeber, L., Armbruster, J.~G., Ozer, N.,
  Michael, A.~J., Baris, S., \& Aktar, M., 2003.
\newblock A shallow fault-zone structure illuminated by trapped waves in the
  karadere--duzce branch of the north anatolian fault, western turkey, {\it
  Geophysical Journal International\/}, {\bf 152}(3), 699--717.

\bibitem[Bhat et~al.(2004)Bhat, Dmowska, Rice, \& Kame]{bhat2004}
Bhat, H.~S., Dmowska, R., Rice, J.~R., \& Kame, N., 2004.
\newblock Dynamic slip transfer from the denali to totschunda faults, alaska:
  Testing theory for fault branching, {\it Bull. Seism. Soc. Am.\/}, {\bf 94},
  S202--S213.

\bibitem[Bhat et~al.(2010)Bhat, Biegel, Rosakis, \& Sammis]{bhat2010a}
Bhat, H.~S., Biegel, R.~L., Rosakis, A.~J., \& Sammis, C.~G., 2010.
\newblock The effect of asymmetric damage on dynamic shear rupture propagation
  ii: With mismatch in bulk elasticity, {\it Tectonophysics\/}, {\bf 493}(3),
  263--271.

\bibitem[Bhat et~al.(2012{\natexlab{a}})Bhat, Rosakis, \& Sammis]{bhat2012}
Bhat, H.~S., Rosakis, A.~J., \& Sammis, C.~G., 2012{\natexlab{a}}.
\newblock A micromechanics based constitutive model for brittle failure at high
  strain rates, {\it J. Appl. Mech.\/}, {\bf 79}(3).

\bibitem[Bhat et~al.(2012{\natexlab{b}})Bhat, Rosakis, \&
  Sammis]{bhat2012micromechanics}
Bhat, H.~S., Rosakis, A.~J., \& Sammis, C.~G., 2012{\natexlab{b}}.
\newblock A micromechanics based constitutive model for brittle failure at high
  strain rates, {\it Journal of Applied Mechanics\/}, {\bf 79}(3).

\bibitem[Biegel \& Sammis(2004)]{Biegel2004}
Biegel, R.~L. \& Sammis, C.~G., 2004.
\newblock {\it Relating Fault Mechanics to Fault Zone Structure\/}, vol.~47,
  pp. 65--111, Elsevier.

\bibitem[Biegel et~al.(2010)Biegel, Bhat, Sammis, \& Rosakis]{biegel2010}
Biegel, R.~L., Bhat, H.~S., Sammis, C.~G., \& Rosakis, A.~J., 2010.
\newblock The effect of asymmetric damage on dynamic shear rupture propagation
  i: No mismatch in bulk elasticity, {\it Tectonophysics\/}, {\bf 493}(3),
  254--262.

\bibitem[Brenguier et~al.(2008)Brenguier, Campillo, Hadziioannou, Shapiro,
  Nadeau, \& Larose]{Brenguier2008}
Brenguier, F., Campillo, M., Hadziioannou, C., Shapiro, N.~M., Nadeau, R.~M.,
  \& Larose, E., 2008.
\newblock Postseismic relaxation along the san andreas fault at parkfield from
  continuous seismological observations, {\it Science\/}, {\bf 321}(5895),
  1478--1481.

\bibitem[Byerlee(1978)]{byerlee1978friction}
Byerlee, J., 1978.
\newblock Friction of rocks, in {\em Rock friction and earthquake
  prediction\/}, pp. 615--626, Springer.

\bibitem[Canitano et~al.(2021)Canitano, Godano, \&
  Thomas]{canitano2021inherited}
Canitano, A., Godano, M., \& Thomas, M.~Y., 2021.
\newblock Inherited state of stress as a key factor controlling slip and slip
  mode: inference from the study of a slow slip event in the longitudinal
  valley, taiwan, {\it Geophysical Research Letters\/}, {\bf 48}(3),
  e2020GL090278.

\bibitem[Cappa et~al.(2014)Cappa, Perrin, Manighetti, \& Delor]{cappa2014off}
Cappa, F., Perrin, C., Manighetti, I., \& Delor, E., 2014.
\newblock Off-fault long-term damage: A condition to account for generic,
  triangular earthquake slip profiles, {\it Geochemistry, Geophysics,
  Geosystems\/}, {\bf 15}(4), 1476--1493.

\bibitem[C{\'e}l{\'e}rier(2008)]{celerier2008seeking}
C{\'e}l{\'e}rier, B., 2008.
\newblock Seeking anderson's faulting in seismicity: a centennial celebration,
  {\it Reviews of Geophysics\/}, {\bf 46}(4).

\bibitem[Chester et~al.(1993)Chester, Evans, \& Biegel]{Chester1993}
Chester, F.~M., Evans, J.~P., \& Biegel, R.~L., 1993.
\newblock Internal structure and weakening mechanisms of the san andreas fault,
  {\it Journal of Geophysical Research-Solid Earth\/}, {\bf 98}(B1), 771--786.

\bibitem[Cochran et~al.(2009)Cochran, Li, Shearer, Barbot, Fialko, \&
  Vidale]{cochran2009seismic}
Cochran, E.~S., Li, Y.-G., Shearer, P.~M., Barbot, S., Fialko, Y., \& Vidale,
  J.~E., 2009.
\newblock Seismic and geodetic evidence for extensive, long-lived fault damage
  zones, {\it Geology\/}, {\bf 37}(4), 315--318.

\bibitem[Collettini \& Sibson(2001)]{Collettini2001}
Collettini, C. \& Sibson, R.~H., 2001.
\newblock Normal faults, normal friction?, {\it Geology\/}, {\bf 29}(10),
  927--930.

\bibitem[Cowie \& Scholz(1992)]{cowie1992growth}
Cowie, P.~A. \& Scholz, C.~H., 1992.
\newblock Growth of faults by accumulation of seismic slip, {\it Journal of
  Geophysical Research: Solid Earth\/}, {\bf 97}(B7), 11085--11095.

\bibitem[Cubas et~al.(2015)Cubas, Lapusta, Avouac, \& Perfettini]{Cubas2015}
Cubas, N., Lapusta, N., Avouac, J.-P., \& Perfettini, H., 2015.
\newblock Numerical modeling of long-term earthquake sequences on the ne japan
  megathrust: Comparison with observations and implications for fault friction,
  {\it Earth and Planetary Science Letters\/}, {\bf 419}, 187--198.

\bibitem[Dalguer et~al.(2003)Dalguer, Irikura, \& Riera]{dalguer2003simulation}
Dalguer, L., Irikura, K., \& Riera, J., 2003.
\newblock Simulation of tensile crack generation by three-dimensional dynamic
  shear rupture propagation during an earthquake, {\it Journal of Geophysical
  Research: Solid Earth\/}, {\bf 108}(B3).

\bibitem[Das \& Aki(1977)]{das1977}
Das, S. \& Aki, K., 1977.
\newblock A numerical study of two-dimensional spontaneous rupture propagation,
  {\it Geophys. J. R. astr. Soc.\/}, {\bf 50}, 643--668.

\bibitem[Das \& Kostrov(1986)]{das1986a}
Das, S. \& Kostrov, B., 1986.
\newblock Fracture of a single asperity on a finite fault: a model for weak
  earthquakes?, in {\em Earthquake Source Mechanics\/}, vol.~6 of {\bf Geophys.
  Monograph}, pp. 91--96.

\bibitem[Das \& Kostrov(1987)]{das1987}
Das, S. \& Kostrov, B.~V., 1987.
\newblock On the numerical boundary integral equation method for
  three-dimensional dynamic shear crack problems, {\it J. Appl. Mech.\/}, {\bf
  54}(1), 99--104.

\bibitem[Day et~al.(2005)Day, Dalguer, Lapusta, \& Liu]{Day2005}
Day, S.~M., Dalguer, L.~A., Lapusta, N., \& Liu, Y., 2005.
\newblock Comparison of finite difference and boundary integral solutions to
  three-dimensional spontaneous rupture, {\it Journal of Geophysical
  Research-solid Earth\/}, {\bf 110}(B12), B12307.

\bibitem[Dor et~al.(2006)Dor, Rockwell, \& Ben-Zion]{Dor2006}
Dor, O., Rockwell, T.~K., \& Ben-Zion, Y., 2006.
\newblock Geological observations of damage asymmetry in the structure of the
  san jacinto, san andreas and punchbowl faults in southern california: A
  possible indicator for preferred rupture propagation direction, {\it pure and
  applied geophysics\/}, {\bf 163}(2), 301--349.

\bibitem[Duan \& Day(2008)]{duan2008inelastic}
Duan, B. \& Day, S.~M., 2008.
\newblock Inelastic strain distribution and seismic radiation from rupture of a
  fault kink, {\it Journal of Geophysical Research: Solid Earth\/}, {\bf
  113}(B12).

\bibitem[Dunham et~al.(2011{\natexlab{a}})Dunham, Belanger, Cong, \&
  Kozdon]{Dunham2011b}
Dunham, E.~M., Belanger, D., Cong, L., \& Kozdon, J.~E., 2011{\natexlab{a}}.
\newblock Earthquake ruptures with strongly rate-weakening friction and
  off-fault plasticity, part 2: Nonplanar faults, {\it Bulletin of the
  Seismological Society of America\/}, {\bf 101}(5), 2308--2322.

\bibitem[Dunham et~al.(2011{\natexlab{b}})Dunham, Belanger, Cong, \&
  Kozdon]{dunham2011earthquake}
Dunham, E.~M., Belanger, D., Cong, L., \& Kozdon, J.~E., 2011{\natexlab{b}}.
\newblock Earthquake ruptures with strongly rate-weakening friction and
  off-fault plasticity, part 1: Planar faults, {\it Bulletin of the
  Seismological Society of America\/}, {\bf 101}(5), 2296--2307.

\bibitem[Erickson et~al.(2017)Erickson, Dunham, \& Khosravifar]{Erickson2017}
Erickson, B.~A., Dunham, E.~M., \& Khosravifar, A., 2017.
\newblock A finite difference method for off-fault plasticity throughout the
  earthquake cycle, {\it Journal of the Mechanics and Physics of Solids\/},
  {\bf 109}, 50--77.

\bibitem[Erickson et~al.(2020)Erickson, Jiang, Barall, Lapusta, Dunham, Harris,
  Abrahams, Allison, Ampuero, Barbot, Cattania, Elbanna, Fialko, Idini, Kozdon,
  Lambert, Liu, Luo, Ma, Best~McKay, Segall, Shi, van~den Ende, \&
  Wei]{Erickson2020}
Erickson, B.~A., Jiang, J., Barall, M., Lapusta, N., Dunham, E.~M., Harris, R.,
  Abrahams, L.~S., Allison, K.~L., Ampuero, J., Barbot, S., Cattania, C.,
  Elbanna, A., Fialko, Y., Idini, B., Kozdon, J.~E., Lambert, V., Liu, Y., Luo,
  Y., Ma, X., Best~McKay, M., Segall, P., Shi, P., van~den Ende, M., \& Wei,
  M., 2020.
\newblock The community code verification exercise for simulating sequences of
  earthquakes and aseismic slip (seas), {\it Seismological Research Letters\/},
  {\bf 91}(2A), 874--890.

\bibitem[Faulkner et~al.(2006)Faulkner, Mitchell, Healy, \&
  Heap]{faulkner2006slip}
Faulkner, D., Mitchell, T., Healy, D., \& Heap, M., 2006.
\newblock Slip on'weak'faults by the rotation of regional stress in the
  fracture damage zone, {\it Nature\/}, {\bf 444}(7121), 922--925.

\bibitem[Faulkner et~al.(2011)Faulkner, Mitchell, Jensen, \&
  Cembrano]{Faulkner2011}
Faulkner, D.~R., Mitchell, T.~M., Jensen, E., \& Cembrano, J., 2011.
\newblock Scaling of fault damage zones with displacement and the implications
  for fault growth processes, {\it J. Geophys. Res.\/}, {\bf 116}(B5).

\bibitem[Fliss et~al.(2005)Fliss, Bhat, Dmowska, \& Rice]{fliss2005}
Fliss, S., Bhat, H.~S., Dmowska, R., \& Rice, J.~R., 2005.
\newblock Fault branching and rupture directivity, {\it J. Geophys. Res.\/},
  {\bf B06312}.

\bibitem[Froment et~al.(2014)Froment, McGuire, van~der Hilst, Gou{\'e}dard,
  Roland, Zhang, \& Collins]{froment2014imaging}
Froment, B., McGuire, J., van~der Hilst, R., Gou{\'e}dard, P., Roland, E.,
  Zhang, H., \& Collins, J., 2014.
\newblock Imaging along-strike variations in mechanical properties of the gofar
  transform fault, east pacific rise, {\it Journal of Geophysical Research:
  Solid Earth\/}, {\bf 119}(9), 7175--7194.

\bibitem[Fukuyama et~al.(2003)Fukuyama, Ellsworth, Waldhauser, \&
  Kubo]{fukuyama2003detailed}
Fukuyama, E., Ellsworth, W.~L., Waldhauser, F., \& Kubo, A., 2003.
\newblock Detailed fault structure of the 2000 western tottori, japan,
  earthquake sequence, {\it Bulletin of the Seismological Society of
  America\/}, {\bf 93}(4), 1468--1478.

\bibitem[Gabriel et~al.(2013)Gabriel, Ampuero, Dalguer, \& Mai]{Gabriel2013}
Gabriel, A.~A., Ampuero, J.-P., Dalguer, L.~A., \& Mai, P.~M., 2013.
\newblock Source properties of dynamic rupture pulses with off-fault
  plasticity, {\it Journal of Geophysical Research-solid Earth\/}, {\bf
  118}(8), 4117--4126.

\bibitem[Healy et~al.(2015)Healy, Blenkinsop, Timms, Meredith, Mitchell, \&
  Cooke]{Healy2015}
Healy, D., Blenkinsop, T.~G., Timms, N.~E., Meredith, P.~G., Mitchell, T.~M.,
  \& Cooke, M.~L., 2015.
\newblock Polymodal faulting: Time for a new angle on shear failure, {\it
  Journal of Structural Geology\/}, {\bf 80}, 57--71.

\bibitem[Hiramatsu et~al.(2005)Hiramatsu, Honma, Saiga, Furumoto, \&
  Ooida]{hiramatsu2005seismological}
Hiramatsu, Y., Honma, H., Saiga, A., Furumoto, M., \& Ooida, T., 2005.
\newblock Seismological evidence on characteristic time of crack healing in the
  shallow crust, {\it Geophysical Research Letters\/}, {\bf 32}(9).

\bibitem[Hok et~al.(2010)Hok, Campillo, Cotton, Favreau, \& Ionescu]{Hok2010}
Hok, S., Campillo, M., Cotton, F., Favreau, P., \& Ionescu, I., 2010.
\newblock Off-fault plasticity favors the arrest of dynamic ruptures on
  strength heterogeneity: Two-dimensional cases, {\it Geophysical Research
  Letters\/}, {\bf 37}, L02306.

\bibitem[Huang et~al.(2014)Huang, Ampuero, \& Helmberger]{Huang2014a}
Huang, Y., Ampuero, J.-P., \& Helmberger, D.~V., 2014.
\newblock Earthquake ruptures modulated by waves in damaged fault zones, {\it
  J. Geophys. Res. Solid Earth\/}, {\bf 119}(4), 3133--3154.

\bibitem[Idini \& Ampuero(2020)]{Idini2020}
Idini, B. \& Ampuero, J.~P., 2020.
\newblock Fault-zone damage promotes pulse-like rupture and back-propagating
  fronts via quasi-static effects, {\it Geophysical Research Letters\/}, {\bf
  47}(23), e2020GL090736.

\bibitem[Im et~al.(2020)Im, Saffer, Marone, \& Avouac]{Im2020}
Im, K., Saffer, D., Marone, C., \& Avouac, J.-P., 2020.
\newblock Slip-rate-dependent friction as a universal mechanism for slow slip
  events, {\it Nature Geoscience\/}, {\bf 13}(10), 705--710.

\bibitem[Jaeger(1979)]{jaeger1979cook}
Jaeger, J., 1979.
\newblock Cook., ngw fundamentals of rock mechanics, {\it Loydon, methuen\/}.

\bibitem[Johnson et~al.(2021)Johnson, Song, Vel, Song, \& Gerbi]{Johnson2021}
Johnson, S.~E., Song, W.~J., Vel, S.~S., Song, B.~R., \& Gerbi, C.~C., 2021.
\newblock Energy partitioning, dynamic fragmentation, and off-fault damage in
  the earthquake source volume, {\it Journal of Geophysical Research: Solid
  Earth\/}, {\bf 126}(11), e2021JB022616, e2021JB022616 2021JB022616.

\bibitem[Johri et~al.(2014)Johri, Dunham, Zoback, \& Fang]{Johri2014}
Johri, M., Dunham, E.~M., Zoback, M.~D., \& Fang, Z., 2014.
\newblock Predicting fault damage zones by modeling dynamic rupture propagation
  and comparison with field observations, {\it Journal of Geophysical Research:
  Solid Earth\/}, {\bf 119}(2), 1251--1272.

\bibitem[Kame et~al.(2003{\natexlab{a}})Kame, Rice, \& Dmowska]{Kame2003}
Kame, N., Rice, J.~R., \& Dmowska, R., 2003{\natexlab{a}}.
\newblock Effects of prestress state and rupture velocity on dynamic fault
  branching, {\it J. Geophys. Res.\/}, {\bf 108}(B5).

\bibitem[Kame et~al.(2003{\natexlab{b}})Kame, Rice, \& Dmowska]{kame2003b}
Kame, N., Rice, J.~R., \& Dmowska, R., 2003{\natexlab{b}}.
\newblock Effects of prestress state and rupture velocity on dynamic fault
  branching, {\it J. Geophys. Res.\/}, {\bf 108}(B5).

\bibitem[Kanamori(2006)]{Kanamori2006}
Kanamori, H., 2006.
\newblock Lessons from the 2004 sumatra-andaman earthquake, {\it Philosophical
  Transactions of the Royal Society of London A: Mathematical, Physical and
  Engineering Sciences\/}, {\bf 364}(1845), 1927--1945.

\bibitem[Kaneko et~al.(2011)Kaneko, Ampuero, \& Lapusta]{Kaneko2011}
Kaneko, Y., Ampuero, J.~P., \& Lapusta, N., 2011.
\newblock Spectral-element simulations of long-term fault slip: Effect of
  low-rigidity layers on earthquake-cycle dynamics, {\it Journal of Geophysical
  Research: Solid Earth\/}, {\bf 116}(B10313).

\bibitem[Kato(2004)]{Kato2004}
Kato, N., 2004.
\newblock Interaction of slip on asperities: Numerical simulation of seismic
  cycles on a two-dimensional planar fault with nonuniform frictional property,
  {\it Journal of Geophysical Research-Solid Earth\/}, {\bf 109}(B12), B12306.

\bibitem[Kelly et~al.(1998)Kelly, Sanderson, \& Peacock]{kelly1998linkage}
Kelly, P., Sanderson, D., \& Peacock, D., 1998.
\newblock Linkage and evolution of conjugate strike-slip fault zones in
  limestones of somerset and northumbria, {\it Journal of Structural
  Geology\/}, {\bf 20}(11), 1477--1493.

\bibitem[King et~al.(1994)King, Stein, \& Lin]{king1994coulomb}
King, G., Stein, R., \& Lin, J., 1994.
\newblock Coulomb static stress technique and the triggering of earthquakes,
  {\it Geophys. J. Int\/}, {\bf 146}, 747--759.

\bibitem[Kostrov(1964)]{kostrov1964}
Kostrov, B.~V., 1964.
\newblock {Selfsimilar problems of propagation of shear cracks}, {\it J. Appl.
  Math. Mech.-USS.\/}, {\bf 28}(5), 1077--1087.

\bibitem[Lapusta et~al.(2000)Lapusta, Rice, Ben-Zion, \&
  Zheng]{lapusta2000elastodynamic}
Lapusta, N., Rice, J.~R., Ben-Zion, Y., \& Zheng, G., 2000.
\newblock Elastodynamic analysis for slow tectonic loading with spontaneous
  rupture episodes on faults with rate-and state-dependent friction, {\it
  Journal of Geophysical Research: Solid Earth\/}, {\bf 105}(B10),
  23765--23789.

\bibitem[Lecomte et~al.(2010)Lecomte, Jolivet, Lacombe, Den\`ele, Labrousse, \&
  Le~Pourhiet]{Lecomte2010}
Lecomte, E., Jolivet, L., Lacombe, O., Den\`ele, Y., Labrousse, L., \&
  Le~Pourhiet, L., 2010.
\newblock Geometry and kinematics of mykonos detachment, cyclades, greece:
  Evidence for slip at shallow dip, {\it Tectonics\/}, {\bf 29}(5), n/a--n/a.

\bibitem[Lisle et~al.(2006)Lisle, Orife, Arlegui, Liesa, \&
  Srivastava]{lisle2006favoured}
Lisle, R.~J., Orife, T.~O., Arlegui, L., Liesa, C., \& Srivastava, D.~C., 2006.
\newblock Favoured states of palaeostress in the earth's crust: evidence from
  fault-slip data, {\it Journal of Structural Geology\/}, {\bf 28}(6),
  1051--1066.

\bibitem[Liu et~al.(2020)Liu, Duan, \& Luo]{Liu2020}
Liu, D., Duan, B., \& Luo, B., 2020.
\newblock Eqsimu: a 3-d finite element dynamic earthquake simulator for
  multicycle dynamics of geometrically complex faults governed by rate- and
  state-dependent friction, {\it Geophysical Journal International\/}, {\bf
  220}(1), 598--609.

\bibitem[Lyakhovsky et~al.(1997)Lyakhovsky, Ben-Zion, \& Agnon]{Lyakhovsky1997}
Lyakhovsky, V., Ben-Zion, Y., \& Agnon, A., 1997.
\newblock Distributed damage, faulting, and friction, {\it Journal of
  Geophysical Research-solid Earth\/}, {\bf 102}(B12), 27635--27649.

\bibitem[Ma(2008)]{Ma2008}
Ma, S., 2008.
\newblock A physical model for widespread near-surface and fault zone damage
  induced by earthquakes, {\it Geochem. Geophys. Geosyst.\/}, {\bf 9}(11).

\bibitem[Madariaga(1976)]{madariaga1976}
Madariaga, R., 1976.
\newblock Dynamics of an expanding circular fault, {\it Bull. Seism. Soc.
  Am.\/}, {\bf 66}(3), 639--666.

\bibitem[Manighetti et~al.(2001)Manighetti, King, Gaudemer, Scholz, \&
  Doubre]{manighetti2001slip}
Manighetti, I., King, G., Gaudemer, Y., Scholz, C., \& Doubre, C., 2001.
\newblock Slip accumulation and lateral propagation of active normal faults in
  afar, {\it Journal of Geophysical Research: Solid Earth\/}, {\bf 106}(B7),
  13667--13696.

\bibitem[Manighetti et~al.(2004)Manighetti, King, \&
  Sammis]{manighetti2004role}
Manighetti, I., King, G., \& Sammis, C.~G., 2004.
\newblock The role of off-fault damage in the evolution of normal faults, {\it
  Earth and Planetary Science Letters\/}, {\bf 217}(3-4), 399--408.

\bibitem[Mitchell \& Faulkner(2009)]{mitchell2009nature}
Mitchell, T. \& Faulkner, D., 2009.
\newblock The nature and origin of off-fault damage surrounding strike-slip
  fault zones with a wide range of displacements: A field study from the
  atacama fault system, northern chile, {\it Journal of Structural Geology\/},
  {\bf 31}(8), 802--816.

\bibitem[Ngo et~al.(2012)Ngo, Huang, Rosakis, Griffith, \& Pollard]{ngo2012off}
Ngo, D., Huang, Y., Rosakis, A., Griffith, W., \& Pollard, D., 2012.
\newblock Off-fault tensile cracks: A link between geological fault
  observations, lab experiments, and dynamic rupture models, {\it Journal of
  Geophysical Research: Solid Earth\/}, {\bf 117}(B1).

\bibitem[Okubo et~al.(2019)Okubo, Bhat, Rougier, Marty, Schubnel, Lei, Knight,
  \& Klinger]{okubo2019dynamics}
Okubo, K., Bhat, H.~S., Rougier, E., Marty, S., Schubnel, A., Lei, Z., Knight,
  E.~E., \& Klinger, Y., 2019.
\newblock Dynamics, radiation, and overall energy budget of earthquake rupture
  with coseismic off-fault damage, {\it Journal of Geophysical Research: Solid
  Earth\/}, {\bf 124}(11), 11771--11801.

\bibitem[Ozawa \& Ando(2021)]{ozawa2021}
Ozawa, S. \& Ando, R., 2021.
\newblock Mainshock and aftershock sequence simulation in geometrically complex
  fault zones, {\it J. Geophys. Res.\/}, {\bf 126}(2).

\bibitem[Palmer \& Rice(1973)]{palmer1973growth}
Palmer, A.~C. \& Rice, J.~R., 1973.
\newblock The growth of slip surfaces in the progressive failure of
  over-consolidated clay, {\it Proceedings of the Royal Society of London. A.
  Mathematical and Physical Sciences\/}, {\bf 332}(1591), 527--548.

\bibitem[Poliakov et~al.(2002)Poliakov, Dmowska, \& Rice]{poliakov2002dynamic}
Poliakov, A.~N., Dmowska, R., \& Rice, J.~R., 2002.
\newblock Dynamic shear rupture interactions with fault bends and off-axis
  secondary faulting, {\it Journal of Geophysical Research: Solid Earth\/},
  {\bf 107}(B11), ESE--6.

\bibitem[Preuss et~al.(2020)Preuss, Ampuero, Gerya, \& van Dinther]{Preuss2020}
Preuss, S., Ampuero, J.~P., Gerya, T.~V., \& van Dinther, Y., 2020.
\newblock Characteristics of earthquake ruptures and dynamic off-fault
  deformation on propagating faults, {\it Solid Earth\/}.

\bibitem[Rice(2002)]{Rice2002}
Rice, J.~R., 2002.
\newblock New perspectives on crack and fault dynamics, in {\em Mechanics for a
  New Millennium: Proceedings of the 20th International Congress of Theoretical
  and Applied Mechanics Chicago, Illinois, USA 27 August - 2 September 2000\/},
  pp. 1--24, Springer Netherlands, Dordrecht.

\bibitem[Rice et~al.(2005)Rice, Sammis, \& Parsons]{rice2005off}
Rice, J.~R., Sammis, C.~G., \& Parsons, R., 2005.
\newblock Off-fault secondary failure induced by a dynamic slip pulse, {\it
  Bulletin of the Seismological Society of America\/}, {\bf 95}(1), 109--134.

\bibitem[Richards-Dinger \& Shearer(2000)]{RichardsDinger2000}
Richards-Dinger, K.~B. \& Shearer, P.~M., 2000.
\newblock Earthquake locations in southern california obtained using
  source-specific station terms, {\it Journal of Geophysical Research-Solid
  Earth\/}, {\bf 105}(B5), 10939--10960.

\bibitem[Rodriguez~Padilla et~al.(2022)Rodriguez~Padilla, Oskin, Milliner, \&
  Plesch]{Rodriguez-Padilla2022}
Rodriguez~Padilla, A.~M., Oskin, M.~E., Milliner, C. W.~D., \& Plesch, A.,
  2022.
\newblock Accrual of widespread rock damage from the 2019 ridgecrest
  earthquakes, {\it Nature Geoscience\/}, {\bf 15}(3), 222--226.

\bibitem[Romanet et~al.(2018)Romanet, Bhat, Jolivet, \&
  Madariaga]{romanet2018fast}
Romanet, P., Bhat, H.~S., Jolivet, R., \& Madariaga, R., 2018.
\newblock Fast and slow slip events emerge due to fault geometrical complexity,
  {\it Geophysical Research Letters\/}, {\bf 45}(10), 4809--4819.

\bibitem[Sammis et~al.(2009)Sammis, Rosakis, \& Bhat]{sammis2009}
Sammis, C.~G., Rosakis, A.~J., \& Bhat, H.~S., 2009.
\newblock Effects of off-fault damage on earthquake rupture propagation:
  Experimental studies, {\it Pure Appl. Geophys.\/}, {\bf 166}.

\bibitem[Savage \& Brodsky(2011)]{Savage2011}
Savage, H.~M. \& Brodsky, E.~E., 2011.
\newblock Collateral damage: Evolution with displacement of fracture
  distribution and secondary fault strands in fault damage zones.

\bibitem[Shi \& Ben-Zion(2006)]{Shi2006}
Shi, Z. \& Ben-Zion, Y., 2006.
\newblock Dynamic rupture on a bimaterial interface governed by slip-weakening
  friction, {\it Geophysical Journal International\/}, {\bf 165}(2), 469--484.

\bibitem[Shipton \& Cowie(2001)]{Shipton2001}
Shipton, Z. \& Cowie, P., 2001.
\newblock Damage zone and slip-surface evolution over $\mu$m to km scales in
  high-porosity navajo sandstone, utah, {\it Journal of Structural Geology\/},
  {\bf 23}(12), 1825--1844.

\bibitem[Sibson et~al.(2012)Sibson, Ghisetti, \&
  Crookbain]{sibson2012andersonian}
Sibson, R., Ghisetti, F., \& Crookbain, R., 2012.
\newblock Andersonian wrench faulting in a regional stress field during the
  2010--2011 canterbury, new zealand, earthquake sequence, {\it Geological
  Society, London, Special Publications\/}, {\bf 367}(1), 7--18.

\bibitem[Stein et~al.(1997)Stein, Barka, \& Dieterich]{stein1997progressive}
Stein, R.~S., Barka, A.~A., \& Dieterich, J.~H., 1997.
\newblock Progressive failure on the north anatolian fault since 1939 by
  earthquake stress triggering, {\it Geophysical Journal International\/}, {\bf
  128}(3), 594--604.

\bibitem[Tal \& Faulkner(2022)]{Tal2022}
Tal, Y. \& Faulkner, D.~R., 2022.
\newblock The effect of fault roughness and earthquake ruptures on the
  evolution and scaling of fault damage zones, {\it Journal of Geophysical
  Research: Solid Earth\/}, {\bf 127}.

\bibitem[Templeton \& Rice(2008)]{templeton2008off}
Templeton, E.~L. \& Rice, J.~R., 2008.
\newblock Off-fault plasticity and earthquake rupture dynamics: 1. dry
  materials or neglect of fluid pressure changes, {\it Journal of Geophysical
  Research: Solid Earth\/}, {\bf 113}(B9).

\bibitem[Thakur \& Huang(2021)]{Thakur2021}
Thakur, P. \& Huang, Y., 2021.
\newblock Influence of fault zone maturity on fully dynamic earthquake cycles,
  {\it Geophysical Research Letters\/}, {\bf 48}(17), e2021GL094679.

\bibitem[Thomas \& Bhat(2018)]{thomas2018dynamic}
Thomas, M.~Y. \& Bhat, H.~S., 2018.
\newblock Dynamic evolution of off-fault medium during an earthquake: a
  micromechanics based model, {\it Geophysical Journal International\/}, {\bf
  214}(2), 1267--1280.

\bibitem[Thomas et~al.(2014)Thomas, Lapusta, Noda, \& Avouac]{thomas2014quasi}
Thomas, M.~Y., Lapusta, N., Noda, H., \& Avouac, J.-P., 2014.
\newblock Quasi-dynamic versus fully dynamic simulations of earthquakes and
  aseismic slip with and without enhanced coseismic weakening, {\it Journal of
  Geophysical Research: Solid Earth\/}, {\bf 119}(3), 1986--2004.

\bibitem[Thomas et~al.(2017{\natexlab{a}})Thomas, Avouac, \&
  Lapusta]{thomas2017rate}
Thomas, M.~Y., Avouac, J.-P., \& Lapusta, N., 2017{\natexlab{a}}.
\newblock Rate-and-state friction properties of the longitudinal valley fault
  from kinematic and dynamic modeling of seismic and aseismic slip, {\it
  Journal of Geophysical Research: Solid Earth\/}, {\bf 122}(4), 3115--3137.

\bibitem[Thomas et~al.(2017{\natexlab{b}})Thomas, Bhat, \&
  Klinger]{thomas2017effect}
Thomas, M.~Y., Bhat, H.~S., \& Klinger, Y., 2017{\natexlab{b}}.
\newblock Effect of brittle off-fault damage on earthquake rupture dynamics,
  {\it Fault zone dynamic processes: Evolution of fault properties during
  seismic rupture\/}, {\bf 227}, 255.

\bibitem[Townend \& Zoback(2000)]{townend2000faulting}
Townend, J. \& Zoback, M.~D., 2000.
\newblock How faulting keeps the crust strong, {\it Geology\/}, {\bf 28}(5),
  399--402.

\bibitem[Tullis \& Schubert(2015)]{Tullis2015}
Tullis, T.~E. \& Schubert, G., 2015.
\newblock 4.06 - mechanisms for friction of rock at earthquake slip rates, in
  {\em Treatise on Geophysics (Second Edition)\/}, pp. 139--159, Elsevier,
  Oxford.

\bibitem[Uphoff et~al.(2022)Uphoff, May, \& Gabriel]{Uphoff2022}
Uphoff, C., May, D.~A., \& Gabriel, A.-A., 2022.
\newblock {A discontinuous Galerkin method for sequences of earthquakes and
  aseismic slip on multiple faults using unstructured curvilinear grids}, {\it
  Geophysical Journal International\/}, ggac467.

\bibitem[van~den Ende et~al.(2018)van~den Ende, Chen, Ampuero, \&
  Niemeijer]{Ende2018}
van~den Ende, M. P.~A., Chen, J., Ampuero, J.~P., \& Niemeijer, A.~R., 2018.
\newblock A comparison between rate-and-state friction and microphysical
  models, based on numerical simulations of fault slip, {\it Tectonophysics\/},
  {\bf 733}, 273--295.

\bibitem[Vermilye \& Scholz(1998)]{Vermilye1998}
Vermilye, J.~M. \& Scholz, C.~H., 1998.
\newblock The process zone: A microstructural view of fault growth, {\it
  Journal of Geophysical Research-solid Earth\/}, {\bf 103}(B6), 12223--12237.

\bibitem[Walsh(1965{\natexlab{a}})]{Walsh1965}
Walsh, J.~B., 1965{\natexlab{a}}.
\newblock The effect of cracks in rocks on poisson's ratio, {\it J. Geophys.
  Res.\/}, {\bf 70}(20), 5249--5257.

\bibitem[Walsh(1965{\natexlab{b}})]{Walsh1965a}
Walsh, J.~B., 1965{\natexlab{b}}.
\newblock The effect of cracks on the compressibility of rock, {\it J. Geophys.
  Res.\/}, {\bf 70}(2), 381--389.

\bibitem[Wilson et~al.(2003)Wilson, Chester, \& Chester]{Wilson2003}
Wilson, J.~E., Chester, J.~S., \& Chester, F.~M., 2003.
\newblock Microfracture analysis of fault growth and wear processes, punchbowl
  fault, san andreas system, california, {\it Journal of Structural Geology\/},
  {\bf 25}(11), 1855--1873.

\bibitem[Xu et~al.(2014)Xu, Ben-Zion, Ampuero, \& Lyakhovsky]{Xu2014}
Xu, S., Ben-Zion, Y., Ampuero, J.-P., \& Lyakhovsky, V., 2014.
\newblock Dynamic ruptures on a frictional interface with off-fault brittle
  damage: Feedback mechanisms and effects on slip and near-fault motion, {\it
  Pure and Applied Geophysics\/}, {\bf 172}(5), 1243--1267.

\bibitem[Yamashita(2000)]{yamashita2000generation}
Yamashita, T., 2000.
\newblock Generation of microcracks by dynamic shear rupture and its effects on
  rupture growth and elastic wave radiation, {\it Geophysical Journal
  International\/}, {\bf 143}(2), 395--406.

\bibitem[Yukutake et~al.(2007)Yukutake, Iio, Katao, \&
  Shibutani]{yukutake2007estimation}
Yukutake, Y., Iio, Y., Katao, H., \& Shibutani, T., 2007.
\newblock Estimation of the stress field in the region of the 2000 western
  tottori earthquake: Using numerous aftershock focal mechanisms, {\it Journal
  of Geophysical Research: Solid Earth\/}, {\bf 112}(B9).

\bibitem[Zoback et~al.(1987)Zoback, Zoback, Mount, Suppe, Eaton, Healy,
  Oppenheimer, Reasenberg, Jones, Raleigh, Wong, Scotti, \&
  Wentworth]{Zoback1987}
Zoback, M.~D., Zoback, M.~L., Mount, V.~S., Suppe, J., Eaton, J.~P., Healy,
  J.~H., Oppenheimer, D., Reasenberg, P., Jones, L., Raleigh, C.~B., Wong,
  I.~G., Scotti, O., \& Wentworth, C., 1987.
\newblock New evidence on the state of stress of the san andreas fault system,
  {\it Science\/}, {\bf 238}(4830), 1105--1111.

\end{thebibliography}
\end{document}